\RequirePackage{rotating}
\documentclass[fleqn,usenatbib]{mnras}
\usepackage{comment}
\usepackage{gensymb}
\usepackage{url}
\usepackage{caption}
\usepackage{newtxtext,newtxmath}
\usepackage{adjustbox}
\newcommand{\HI}{H\,{\sc i}}
\newcommand{\FHI}{$F_{\rm HI}$}
\newcommand{\kms}{~km\,s$^{-1}$}
\usepackage[autostyle]{csquotes}
\usepackage{textcomp}
\usepackage{rotating}
\usepackage{float}
\MakeRobust{\overrightarrow}

\usepackage{hyperref}
\usepackage[bigfiles]{media9}

\usepackage[T1]{fontenc}
\usepackage{ae,aecompl}
\usepackage{rotating}

\usepackage{graphicx}	
\usepackage{amsmath}	
\usepackage{adjustbox}
\usepackage{xcolor}
\usepackage{ulem}

\title[Tidal debris around the NGC~7232/3 triplet]{MeerKAT-64 discovers wide-spread tidal debris in the nearby NGC~7232 galaxy group}
\author[Namumba et al.]
{B.\ Namumba,$^{1}$\thanks{E-mail: bnamumba@gmail.com}
B.\ S.\ Koribalski,$^{2,3}$
G.\ I.\ G.\ J\'ozsa,$^{4,1}$ 
K.\ Lee-Waddell,$^{2}$
M.\ G.\ Jones,$^{5,16}$
\newauthor
C.\ Carignan,$^{6,7,8}$ 
L.\ Verdes-Montenegro,$^{5}$
R.\ Ianjamasimanana,$^{1,4}$ 
W.\ J.\ G.\ de Blok,$^{9,6,10}$
\newauthor
M.\ Cluver,$^{11}$ 
J.\ Garrido,$^{5}$
S.\ S\'{a}nchez-Exp\'{o}sito$^{5}$
A.\ J.\ T.\ Ramaila,$^{4}$
K.\ Thorat,$^{12}$
\newauthor
L.\ A. L. Andati,$^{1}$
B.\ V. Hugo,$^{1,4}$
D.\ Kleiner,$^{13}$
P.\ Kamphuis,$^{14}$
P.\ Serra,$^{13}$ 
\newauthor
O.\ M.\ Smirnov,$^{1,4}$
F.\ M.\ Maccagni,$^{13}$
S.\ Makhathini,$^{17}$
D.\ Cs. Moln\'{a}r,$^{13}$
\newauthor
S.\ Perkins,$^{4}$
M.\ Ramatsoku,$^{1,13}$
S.\ V.\ White,$^{1}$
F.\ Loi,$^{13}$
\\
$^1$Department of Physics and Electronics, Rhodes University, PO Box 94, Makhanda, 6140, South Africa \\ 
$^2$Australia Telescope National Facility, CSIRO Astronomy and Space Science, P.O. Box 76, NSW 1710, Epping, Australia \\
$^3$Western Sydney University, Locked Bag 1797, Penrith, NSW 2751, Australia\\
$^{4}$South African Radio Astronomy Observatory, Black River Park, 2 Fir Street, Observatory, Cape Town, 7925, South Africa\\
$^{5}$Instituto de Astrofísica de Andalucía (CSIC), Glorieta de la Astronomía, 18008 Granada, Spain\\
$^{6}$Department of Astronomy, University of Cape Town, Private Bag X3, Rondebosch 7701, South Africa\\
$^{7}$Département de physique, Université de Montréal, C.P. 6128, Succ. centre-ville, Montréal, Québec, Canada, H3C 3J7\\
$^{8}$Observatoire d$^{\prime}$Astrophysique de l$^{\prime}$Universit$\acute{e}$ de Ouagadougou, BP 7021, Ouagadougou 03, Burkina Faso \\
$^{9}$Netherlands Institute for Radio Astronomy (ASTRON), Oude Hoogeveensedijk 4, 7991 PD Dwingeloo, The Netherlands\\
$^{10}$Kapteyn Astronomical Institute, University of Groningen, Postbus 800, 9700 AV Groningen, The Netherlands\\
$^{11}$Centre for Astrophysics and Supercomputing, Swinburne University of Technology, Hawthorn, Victoria 3122, Australia\\
$^{12}$Department of Physics, University of Pretoria, Hatfield, Pretoria, 0028, South Africa\\
$^{13}$INAF - Osservatorio Astronomico di Cagliari, Via della Scienza 5, I-09047 Selargius (CA), Italy\\
$^{14}$Ruhr University Bochum, Faculty of Physics and Astronomy, Astronomical Institute, 44780 Bochum, Germany \\
$^{15}$Department of Physics and Astronomy, University of the Western Cape, Robert Sobkwe Road, Bellville, 7535, South Africa \\
$^{16}$Steward Observatory, University of Arizona, 933 North Cherry Avenue, Rm. N204, Tucson, AZ 85721-0065, USA\\
$^{17}$ School of Physics, University of the Witwatersrand, 1 Jan Smuts Avenue, Johannesburg, South Africa
}

\date{Accepted 2021 May 19. Received 2021 May 18; in original form 2021 February 4}

\pubyear{2020}

\begin{document}
\label{firstpage}
\pagerange{\pageref{firstpage}--\pageref{lastpage}}
\maketitle

\begin{abstract}
We report the discovery of large amounts of previously undetected cold neutral atomic hydrogen (\HI) around the core triplet galaxies in the nearby NGC~7232 galaxy group with MeerKAT. With a physical resolution of $\sim$1 kpc, we detect a complex web of low surface brightness \HI\ emission down to a 4$\sigma$ column density level of $\sim$1 $\times$ 10$^{19}$ cm$^{-2}$ (over 44 \kms ). The newly discovered H\,{\sc i} streams extend over $\sim$20 arcmin corresponding to 140~kpc in projection. This is $\sim$3 times the \HI\ extent of the galaxy triplet (52 kpc). The \HI\ debris has an \HI\ mass of $\sim$6.6 $\times 10^9$~M$_{\odot}$, more than 50\% of the total \HI\ mass of the triplet. Within the galaxy triplet, NGC~7233 and NGC~7232 have lost a significant amount of \HI\ while NGC~7232B appears to have an excess of \HI. The \HI\ deficiency in NGC~7232 and NGC~7233 indicates that galaxy-galaxy interaction in the group concentrates on this galaxy pair while the other disc galaxies have visited them over time. In comparison to the AMIGA sample of isolated galaxies we find that with regards to its total \HI\ mass the NGC~7232/3 galaxy triplet is not \HI\ deficient. Despite the many interactions associated to the triplet galaxies, no \HI\ seems to have been lost from the group (yet).

\end{abstract}



\section{Introduction}
It is widely known that galaxy properties such as morphology, gas content and star formation are affected by interactions between galaxies and their environments \citep{1980ApJ...236..351D}. As the majority of galaxies in the nearby universe reside in small galaxy groups (e.g.\citealt{2011MNRAS.416.2640R}), it is important to understand how galaxies evolve in such environments \citep{1987ApJ...321..280T,2004MNRAS.348.1355B,2015ApJ...800...24K,2016A&A...596A..14S}. Observational signatures of interaction histories are seen in the form of, e.g., the length and shape of tidal tails, bridges and other debris (e.g.,  \citealt{2004MNRAS.348.1255K,2010AJ....139..102E,2018MNRAS.473.3358S, 2020arXiv200207312K}).

The role of galaxy groups has long been prescribed for understanding galaxy properties over a broad range of environments. In the hierarchical scenario of large-scale structure
formation, galaxy groups and the galaxies therein are  the building blocks of rich clusters \citep{2006ApJ...640..762K}. There is evidence that galaxies must undergo some
amount of pre- processing (e.g, the change in their gas content and morphology) through tidal interaction before falling into clusters (e.g., \citealt{Zabludoff_1998,10.1111/j.1365-2966.2008.13388.x,10.1093/mnras/staa1779}). This indicates that some of the properties of the galaxies  that we see in galaxy clusters today are the results of transformation processes pre-dating the infall into clusters and hence taking place under very different conditions \citep{2004ogci.conf..341D}. As a consequence, galaxy groups are being observed to probe how and by which mechanisms galaxies are pre-processed before they arrive in clusters.

Neutral atomic hydrogen (\HI) is known to be an ideal tracer of current as well as past galaxy interactions. This is because the extended \HI\ envelope, which is typically a factor of 2--3 larger than the observed bright stellar body \citep{2000AJ....120..763P,2012ApJ...756..183B,2018MNRAS.478.1611K} is more susceptible than the optical disc to external disturbances via gravitational and hydrodynamical interactions \citep{1972ApJ...176....1G, 10.1046/j.1365-8711.2001.04102.x, 10.1093/mnras/stw1162, 10.1093/mnras/sty247}.

The effect of the environment on the \HI\ content of galaxies is well studied in dense environments. It is well known that in high-density environment \textcolor{blue}{a} substantial amount of gas goes into the inter-cluster medium (ICM), making the spiral galaxies in rich clusters to be \HI\ deficient i.e., contain less \HI\ mass than field spirals of similar \HI\ morphology and size (e.g., \citealt{2001A&A...376...98S,Solanes_2001,2009AJ....138.1741C}). However, \HI\ deficient galaxies have also been found in compact groups (e.g., \citealt{2001A&A...377..812V,2017MNRAS.464..957H,2019A&A...632A..78J}). Over the past few years, \HI\ deficient galaxies have also been reported in loose groups (e.g., \citealt{2005MNRAS.356...77K,10.1111/j.1365-2966.2006.10307.x,2007MNRAS.378..137S,2013AJ....146..124H}). Of particular interest are hence high-resolution studies of the neutral gas in galaxy groups, to study in detail the transformation processes that lead to its removal from galaxies in groups. Such observations used to be difficult in the past time, often requiring a significant amount of observing time, owing to the limited FOV and sensitivity of suitable telescopes. The number of studies of the resolved properties of \HI\ in groups is therefore limited. Due to these restrictions, one particular question which is poorly addressed in the literature is whether and in which form low-column-density gas in groups exists and what its likely fate is. 

Previous surveys of the \HI\ content in nearby groups have often revealed the presence of intra-group material (e.g., \citealt{2001ApJ...555..232R,2005MNRAS.357L..21B,2005MNRAS.363L..21B}). These observations were typically conducted with large single-dish telescopes  (e.g. HIPASS, \citealt{10.1046/j.1365-8711.2001.04102.x}, and ALFALFA, \citealt{2018ApJ...861...49H}). While single-dish telescopes provide good sensitivity for observing a wide area of the sky in a reasonable amount of time, the coarse spatial resolution of these instruments does not allow to explore the low column density universe on small spatial scales, thus the need for complementary deep, high-angular resolution \HI\ observations. MeerKAT with its high sensitivity and high spatial resolution capabilities enables us to better understand how the low column density phase of \HI\ on small spatial scales relates to galaxy evolution (e.g.,\citealt{2021arXiv210110347K}).

\begin{figure*}
\centering
   \includegraphics[width=12cm]{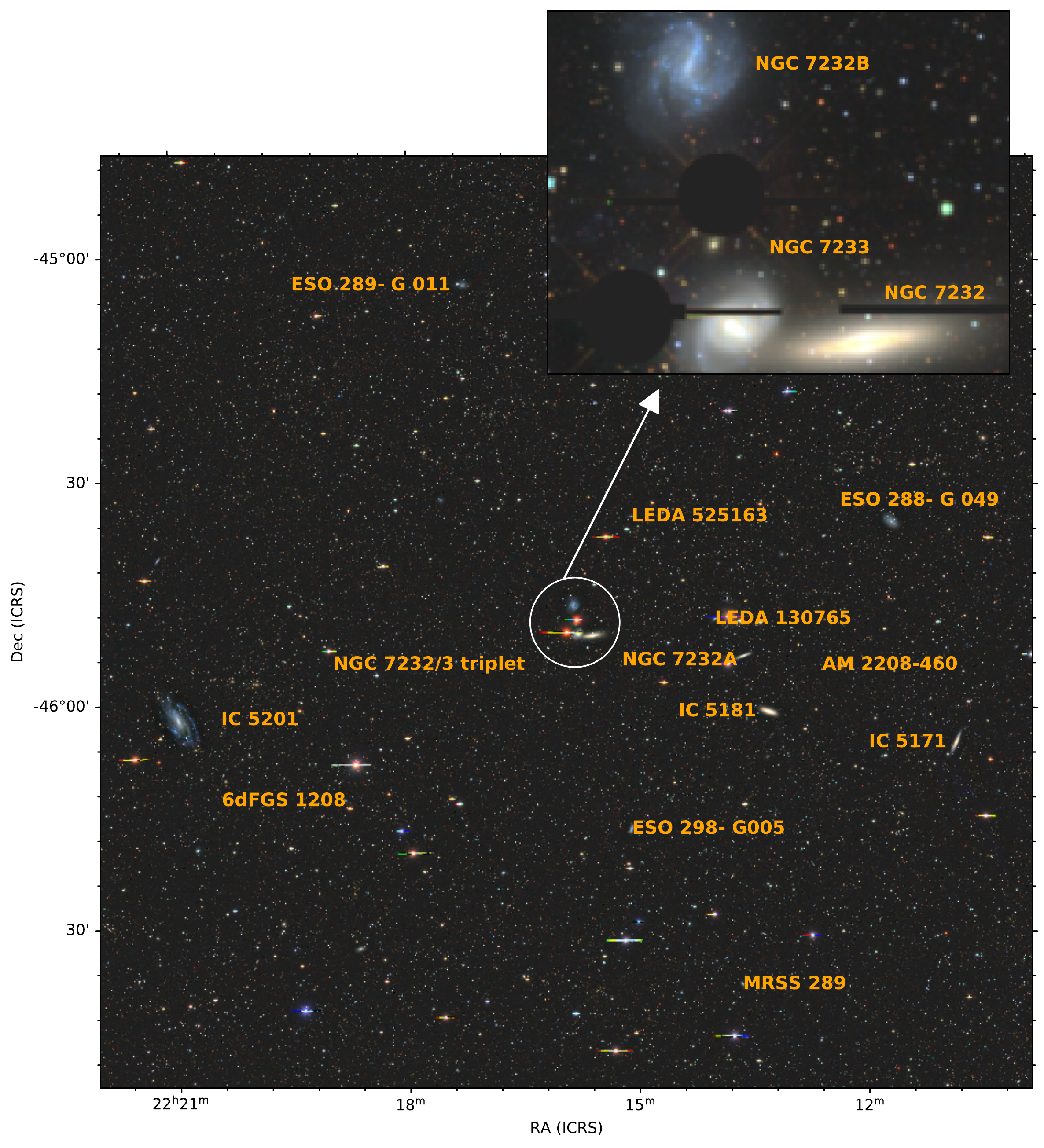}
\caption{Three-colour ($grz$) optical image of the NGC~7232 group from the Dark Energy Legacy Survey with the MeerKAT \HI\ detected galaxies labelled. The white circle indicates the NGC~7232/3 galaxy triplet. The top panel shows the NGC~7232/3 galaxy triplet with bright foreground stars and related artifacts obscured.}
\label{fig:fulloptical}
\end{figure*}

The nearby NGC~7232 galaxy group is located at a distance of $\sim$24~Mpc \citep{1995A&A...297...56G}, which we adopt throughout this paper. This field displays a diverse array of gaseous features, apparent galaxy interactions, as well as gas-rich and gas-poor galaxies, thus making it an excellent target to study how such interactions may impact galaxy evolution. Figure~\ref{fig:fulloptical} shows a $grz$ optical image of the NGC~7232 galaxy group obtained from the Dark Energy Legacy Survey \footnote{www.legacysurvey.org} \citep{2019AJ....157..168D}. Using the \HI\ Parkes All Sky Survey (HIPASS) we estimate that the group has 19 members \citep{2004AJ....128...16K}, four of which are bright galaxies. The three spirals, NGC~7232, NGC~7232B and NGC~7233 lying within 5 arcmin in projection ($\sim$35 kpc) of each other make up the NGC~7232/3 triplet while the lenticular galaxy IC\,5181 lies 10 arcmin southwest of the triplet.

The first interferometric \HI\ studies of the NGC~7232 galaxy group  \citep{10.1046/j.1365-8711.2001.04273.x} were carried out with the Australia Telescope Compact Array (ATCA). The ATCA \HI\ intensity distribution revealed traces of recent interactions within the group, including H\,{\sc i} streams connecting the triplet galaxies. Single-dish HIPASS observations of the NGC~7232 galaxy group show the triplet galaxies are embedded in a common H\,{\sc i} envelope \citep{2004MNRAS.348.1255K}. However, the Parkes observations do not have the spatial resolution to separate these galaxies. Neighbouring galaxies around the galaxy group were also reported. Recently, \citet{2019MNRAS.487.5248L} obtained and analysed ASKAP-12 Early Science data of the NGC~7232 galaxy group. These observations were conducted as part of the Wide-field ASKAP L-Band Legacy All-sky Blind surveY (WALLABY; \citealt{2020Ap&SS.365..118K}) Early Science program. From these observations, 17 \HI\ sources were reported, 5 of which were new galaxies while the other 6 detections appear to be tidal debris in the form of H\,{\sc i} clouds that are associated with the triplet NGC~7232/3 and the S0 galaxy IC\,5181. 

While NGC 7232 is a well-studied group, neither of the previous two interferometric \HI\ studies (\citealt{10.1046/j.1365-8711.2001.04273.x,2019MNRAS.487.5248L}) were able to recover all the \HI\ emission detected by HIPASS  \citep{2004AJ....128...16K}, leaving an open question as to where the missing gas resides. With column density sensitivities of $\sim$ 10$^{20}$ cm$^{-2}$, recent ASKAP observations \citep{2019MNRAS.487.5248L} indicated the presence of undetected \HI\ gas which could form a bridge connecting the galaxy triplet NGC~7232/3 to the surrounding \HI\ clouds. It is likely that an increase in sensitivity could enable the detection of more low column density \HI\ gas and provide an understanding to where the gas resides within the galaxy triplet. This motivated our deeper MeerKAT observations.
We observed the nearby NGC~7232 galaxy group at 1.4 GHz with MeerKAT using the 4k mode (44~km\,s$^{-1}$). At comparable velocity resolution, the sensitivity of MeerKAT allows us to reach a 5 $\times$ lower rms compared to the published ASKAP data \citep{2019MNRAS.487.5248L}. Although the coarse velocity resolution of our observations does not allow us to carry out a detailed analysis of the \HI\ in and around the galaxy triplet, the sensitivity of MeerKAT spectral line observations has allowed us to detect spectacular new \HI\ debris at column densities of $\sim10^{19}$ cm$^{-2}$. The main focus of this paper is to study the \HI\ distribution of the newly detected wide-spread \HI\ debris around the NGC~7232/3 triplet (shown in the white circle, Figure~\ref{fig:fulloptical}) with the aim of understanding the interaction processes between the galaxy triplet with its surrounding environment.

The paper is structured as follows. In Section \ref{sec:datareduction} we describe the MeerKAT observations and data reduction. Section \ref {sec:results} presents the results of the NGC~7232/3 galaxy triplet and its surroundings. Section \ref {sec:discussion} and Section \ref{sec:summary} present our discussion and summary, respectively. 

\section{MeerKAT observations and data reduction}\label{sec:datareduction}
The H\,{\sc i} observations of the galaxy group NGC~7232 were obtained with the MeerKAT telescope \citep{2018ApJ...856..180C} between May 2019 and June 2019. The data were obtained in 3 epochs using a single pointing centered on the NGC~7232/3 triplet ($\alpha,\delta$(J2000) = 22:15:38.4, --45:51:00.3). A total of 61 antennas were used, providing an excellent $uv$ coverage. The observations were conducted in L-band (900 to 1670 MHz) using the SKARAB correlator in 4k-wideband mode (8s integration time), centered at a frequency of 1283.8 MHz. In this mode the band is divided into 4096 channels with a channel width of 209 kHz ($\sim$ 44 km\,s$^{-1}$ for \ion{H}{i} at $z=0$). We observed the bright radio source PKS B1934--6342 as the standard flux and bandpass calibrator for 10 min every $\sim$ 2 h and PKS 2214--3835 as a phase calibrator for 2 min every $\sim$ 15 min. The total integration time on source was 15 h.

The data were reduced using the \textsc{CARACal}\footnote{https://caracal.readthedocs.io/en/latest/} pipeline (\citealt{2020ascl.soft06014J,Jozsa2020}; Makhathini et al., in prep.).
\textsc{CARACal} is an automated H\,{\sc i} and radio continuum data reduction and analysis pipeline. The pipeline runs several open source radio interferometry software packages in containerized environments provided by \textsc{Stimela}\footnote{https://github.com/ratt-ru/Stimela/wiki}, a python-based scripting framework. All data reduction software described below was used in a containerized version inside this framework. We processed a bandwidth of 30 MHz, covering a frequency range between 1400 to 1430 MHz (recession velocities between 4244 to -1930 km\,s$^{-1}$), centred around the H\,{\sc i} line of the NGC 7232 group.

We first flagged bad data due to radio frequency interference (RFI) with \textsc{AOFlagger} \citep{10.1093/mnras/stx1547}. For calibrators, channels with Galactic \HI\ emission were discarded (1417 to 1421 MHz). The corrections to the flux, delay terms, bandpass shapes and antenna gains were determined using the \textsc{CASA} \citep{2007ASPC..376..127M} task \textit{bandpass} while the time-varying phases and antenna gains were corrected using the \textsc{CASA} task \textit{gaincal}. Calibration solutions were then applied to the target and the corrected target visibilities were split from the calibrators using the \textsc{CASA} task \textit{mstransform}. The corrected target visibilities were then flagged with \textsc{AOFlagger}. To improve the data quality, phase-only antenna gain corrections were derived with \textsc{Cubical} \citep{2018MNRAS.478.2399K}, in a self-calibration loop in combination with \textsc{WSclean} \citep{2014MNRAS.444..606O}. Solutions were derived for every 128 s of data.

Imaging of the calibrated data was performed using \textsc{WSclean}. Before spectral line imaging, we replaced the corrected data column with the difference between the corrected data and the model data. This was to subtract the continuum clean model generated from the self-calibration process. An additional continuum subtraction was performed by fitting a 3\textsuperscript{rd} order polynomial to line-free channels using the \textit{uvlin} option in the \textsc{CASA} task \textit{mstransform}. Having separated the H\,{\sc i} emission from the continuum, a data cube was generated using a Robust weighting of 0. For the deconvolution we used a threshold of 0.5$\sigma$ within CLEAN regions created with the Source Finding Application (\textsc{SoFiA}, \citealt{10.1093/mnras/stv079}). The restoring beam of the cube was 10.4$^{\prime \prime} \times 7.2^{\prime \prime}$. To explore the low column density H\,{\sc i} emission, a Gaussian taper of 30$^{\prime \prime}$ was applied to produce a second cube with a synthesized beam of 41.9$^{\prime \prime} \times 35.5 ^{\prime \prime}$ (4.8 kpc $\times$ 4.1 kpc) and rms noise of 0.1 mJy\,beam$^{-1}$. This corresponds to a 4$\sigma$ column density sensitivity of 1 $\times$ 10$^{19}$ cm$^{-2}$ over the velocity range of 44 km\, s$^{-1}$ (one channel). For the analysis presented in this paper, we use the cube produced with a Gaussian taper of 30$^{\prime \prime}$. For comparison, WALLABY reported an rms noise of $\sim$1.6 mJy\,beam$^{-1}$ per 4 km\,s$^{-1}$ channel (or 0.5 mJy\,beam$^{-1}$ for 44 km\,s$^{-1}$) and 30$''$ resolution \citep{2012PASA...29..359K, 2020Ap&SS.365..118K}.

The \HI\ intensity and velocity field maps were created using the 3D source finding application, SoFiA. The smooth and clip method with a signal to noise cutoff of 4$\sigma$ was applied. We allowed SoFiA to independently estimate the noise in each channel by fitting a half-Gaussian distribution to the negative pixels. The reliability threshold was set to 100$\%$. We used two spatial smoothing (Gaussian) kernels of one and two times the beam width over one channel. No smoothing was applied in velocity space because of the low velocity resolution (44 km\,s$^{-1}$) of the data. We checked the setup by independently creating source masks and moment maps based on a 4$\sigma$ cutoff in each channel employing \textsc{CASA} and \textsc{MIRIAD}, with a similar outcome.
\begin{table}
\caption{Parameters of the MeerKAT observations.}
\begin{tabular}{cc}
\hline   
Parameter & NGC~7232 group \\
\hline \hline  
Start of observations & May 2019 \\
End of observations & July 2019\\
Number of antennas & 60 to 64 \\
Baseline range & 29 m to 8 km \\
Total integration & 15 h on source \\
FWHM of primary beam & $\sim$1 $^{\circ}$\\
Channel width & 209 kHz (44 km\,s$^{-1}$)\\
Number of channels & 4096 \\
Flux/bandpass calibrator & PKS B1934--6342\\
Phase calibrator & PKS 2214--3835 \\
\hline  
\multicolumn{2}{c}{Robust = 0 weighting function tapered to 30$^{\prime \prime}$}\\ \\
FWHM of synthesized beam & $41.9^{\prime \prime} \times\ 35.5^{\prime \prime}$ (4.8 kpc $\times$ 4.1 kpc)  \\
RMS noise & 0.1 mJy\\
\HI\ column density limit & \\
(4$\sigma$ over 1 channel = 44 km\,s$^{-1}$) & $\sim1 \times\ 10^{19}$ cm$^{-2}$ \\
\hline    
\end{tabular}
\label{coords_table}
\end{table}  
\section{Results}\label{sec:results}
\subsection{H\,{\sc i} Distribution of the NGC~7232/3 triplet}
\begin{figure*}
\centering
   \includegraphics[width=16cm]{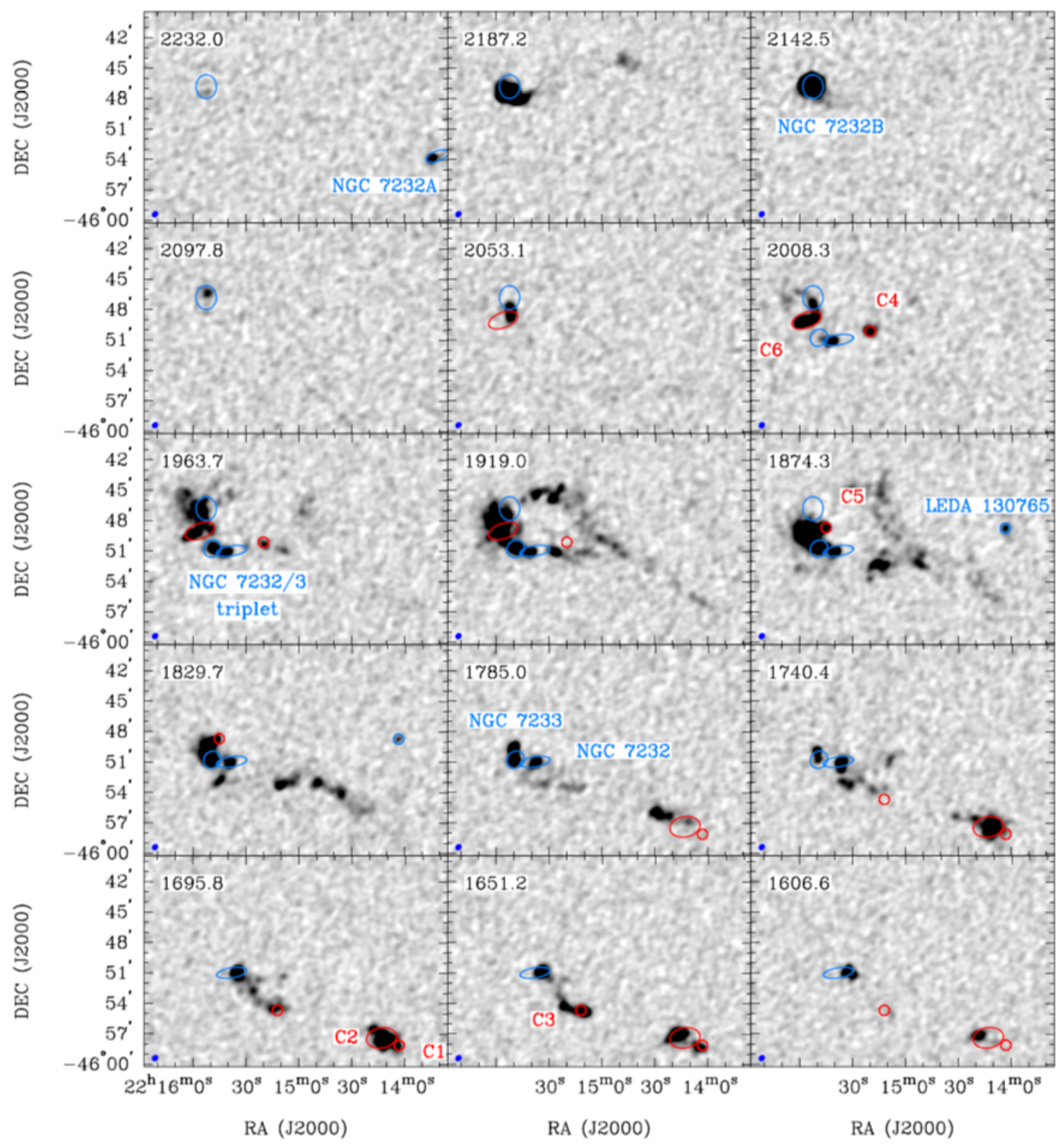}
\caption{MeerKAT-64 H\,{\sc i} channel maps (primary beam corrected) of the NGC~7232/3 triplet, surrounding galaxies, extended tidal streams and debris. Known galaxies are highlighted by blue symbols and labels, while H\,{\sc i} clouds previously detected with ASKAP by \citep{2019MNRAS.487.5248L} are marked in red. The velocities (in km\,s$^{-1}$) are shown in the top left of the single panels and the synthesized beam ($42.0'' \times 35.5''$) in the bottom left.}
\label{fig:HIchannels}
\end{figure*}

The main result from our observations is the detection of wide-spread H\,{\sc i} emission around the galaxy triplet. Only the bright H\,{\sc i} cloud complex between the NGC~7232/3 triplet and IC\,5181 was formerly catalogued as HIPASS J2214--45 \citep{2004AJ....128...16K}. Figure ~\ref{fig:HIchannels} shows the H\,{\sc i} channel maps of the galaxy triplet over the velocity range 1606 -- 2232 km\,s$^{-1}$. While the channel width 44 km\,s$^{-1}$ is coarse, it allows us to trace the different features associated with the NGC~7232/3 triplet. H\,{\sc i} detections already discussed in \cite{2004AJ....128...16K}, \cite{2001MNRAS.324..859B} and \cite{2019MNRAS.487.5248L} are shown in blue and red, while all the unlabelled H\,{\sc i} emission shown at different velocities is detected for the first time, with MeerKAT-64. 
The H\,{\sc i} distribution around the galaxy triplet is complex. Apart from the low column density \HI\ detected around the galaxy \textcolor{blue}{triplet} in all channels, we see a prominent \HI\ tail between velocities of 1829 -- 1919 km\,s$^{-1}$. The \HI\ tail stretches towards the lenticular galaxy IC\,5181, which itself is not detected in H\,{\sc i}. The H\,{\sc i} tail has a length of $\sim$ 13.5$^{\prime}$ (94 kpc at 24 Mpc) in projection. We confirm 6 \HI\ clouds (C1 - C6) which were first detected with ASKAP \citep{2019MNRAS.487.5248L}.

\begin{figure*}
\centering
   \includegraphics[height=12cm]{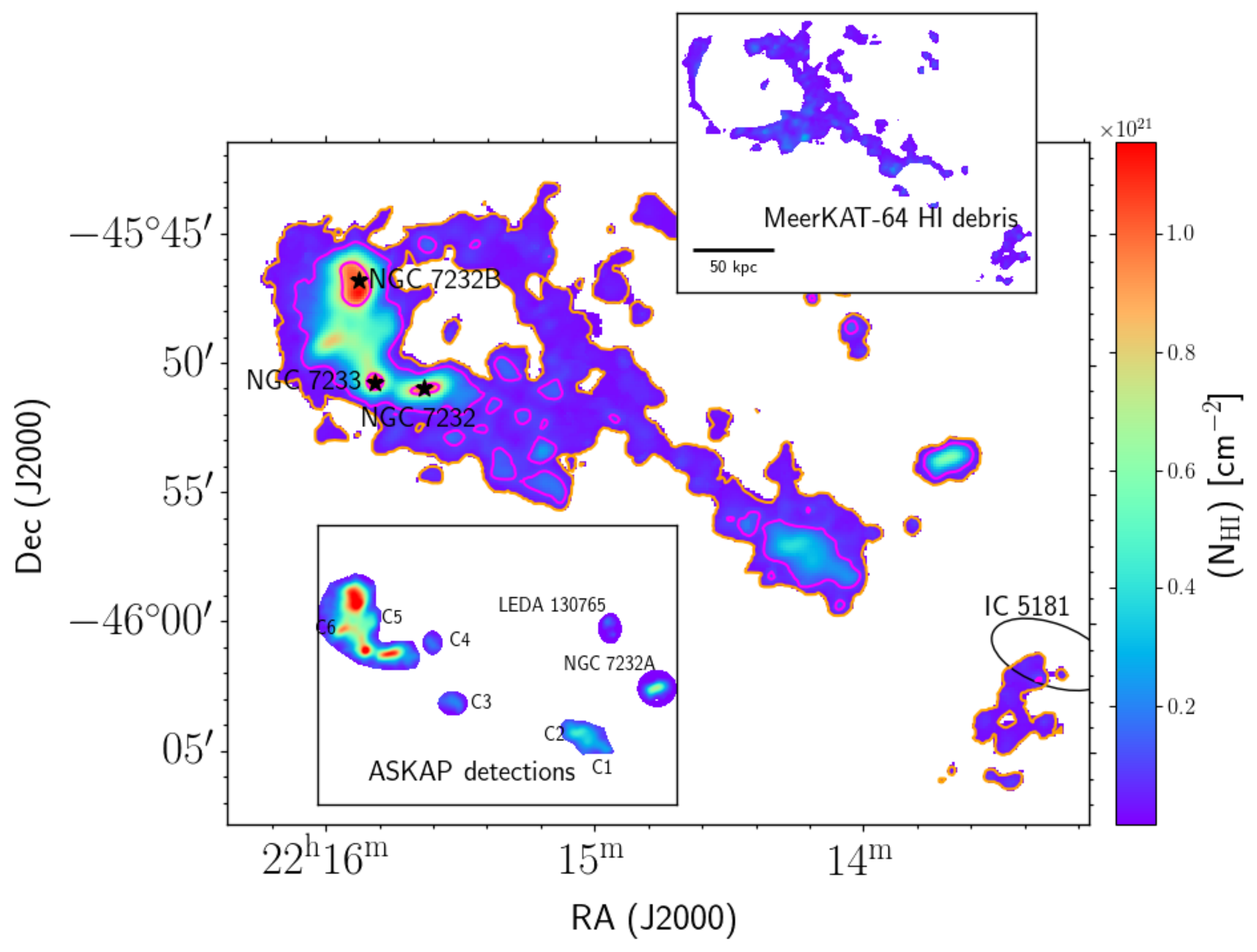}
\caption{MeerKAT-64 H\,{\sc i} column density map (primary beam corrected) of the NGC~7232/3 triplet and its surrounding environment over a velocity range of $\sim$ 1606 to 2232 km\,s$^{-1}$. The black stars indicate the optical centers of the triplet galaxies. Known galaxies and previously identified H\,{\sc i} clouds \citep{2019MNRAS.487.5248L} are shown in the {\bf bottom left inset panel}. The lenticular galaxy, IC\,5181 ($v_{\rm opt}$ = 1987 km\,s$^{-1}$), marked in an ellipse, is not detected in \HI. --- {\bf Main panel:} The \HI\ column density contour levels are 4$\sigma$ $\times$ (1, 16, 80), where the 4$\sigma$ column density limit is 1 $\times$ 10$^{19}$ cm$^{-2}$. The colours represent \HI\ column density levels of $\geq$ 10$^{20}$ cm$^{-2}$ (magenta) and 1 $\times$ 10$^{19}$ cm$^{-2}$ (orange). The MeerKAT detected H\,{\sc i} debris are shown in the {\bf top right inset panel}.}
\label{fig:triplet-mom0}
\end{figure*}

\begin{figure*}
\centering
   \includemedia[3Dtoolbar,3Dmenu,
3Dcoo=-26139.638671875 -441087.40625 -20950.2265625,
3Dc2c=0.0586162731051445 0.9981787204742432 -0.014257380738854408,
3Droo=923195.3480645361,3Droll=-6.457322313032585,3Dlights=Headlamp,3Dlights=Night,transparent=false,height=10cm,activate=onclick]{\includegraphics{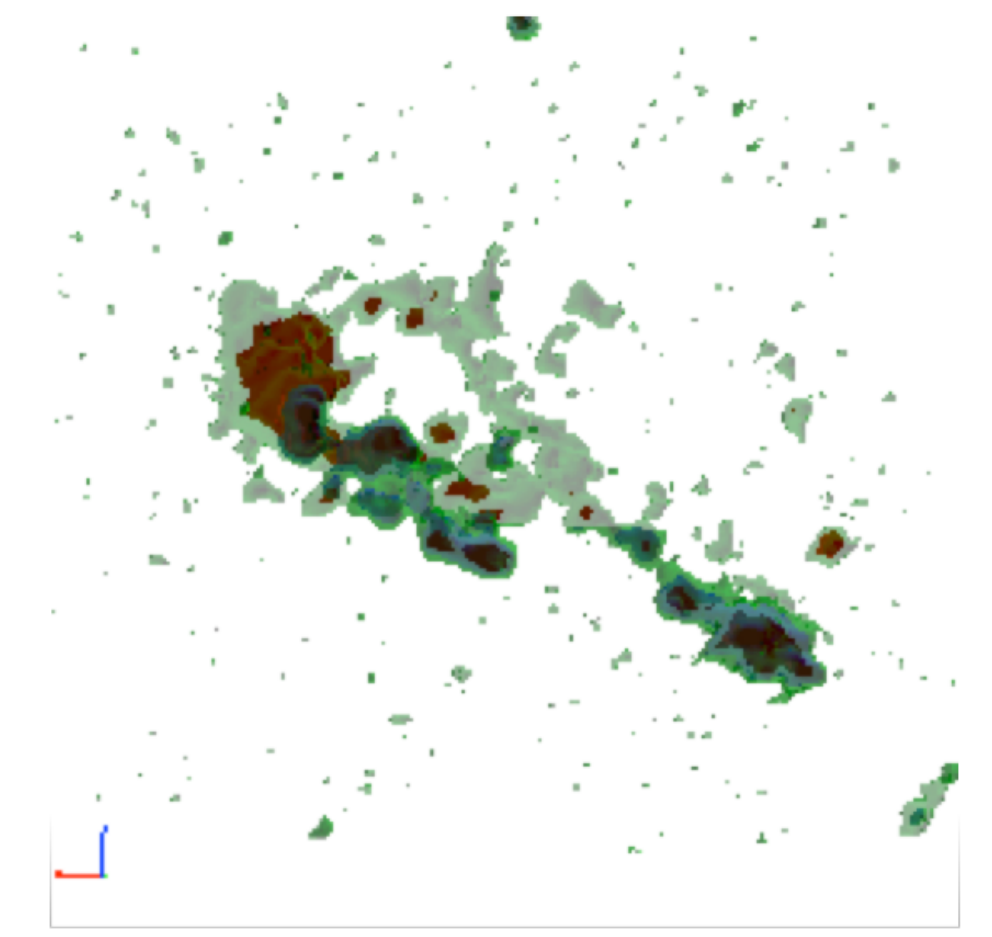}}{ngc7232.u3d}
\caption{3D representation of the \HI\ around the galaxy triplet and its surrounding environment. The green components represent the low column density \HI\ emission, blue colour show the intermediate column density structures, and the red components represent the high column density regions. This must be viewed using Adobe Acrobat.}
\label{fig:3dfinal}
\end{figure*}

Figure~\ref{fig:triplet-mom0} shows the column density map of the galaxy triplet and its surroundings. The bottom-left inset panel shows \HI\ emission at the column density sensitivity of the ASKAP data $\geq 10^{20}$ cm$^{-2}$ while the top inset panel shows the newly detected \HI\ debris with MeerKAT detected at column densities of $\leq 10^{20}$ cm$^{-2}$. Figure~\ref{fig:3dfinal} shows an interactive 3D representation of the galaxy triplet and its surrounding environment created using the X3D pathway introduced in \citet{2016ApJ...818..115V}. The 3D rendering highlights different components of the system and how these components are connected in both position and velocity. The green color represent the low column density \HI\ debris while the blue color show intermediate column density \HI\ structures. The high density \HI\ features are shown in red. Using the primary-beam corrected H\,{\sc i} intensity map, we measure a total flux of F$_{\rm HI}$ = (51.01 $\pm$ 0.29) Jy~km\,s$^{-1}$ and derive a total \HI\ mass of (6.91 $\pm$ 0.03) $\times 10^{9}$ M$_{\odot}$ for the triplet and its surrounding tidal debris. The face-on spiral galaxy NGC~7232B, which is the northern member of the NGC~7232/3 triplet has a total \ion{H}{i} flux of \FHI\ = (9.45 $\pm$ 0.31) Jy~km\,s$^{-1}$ and an \HI\ mass of (1.28 $\pm$ 0.04) $\times 10^{9}$ M$_{\odot}$. We estimate \HI\ fluxes of 
(4.71 $\pm$ 0.26) and (3.32 $\pm$ 0.34) Jy~km\,s$^{-1}$ for the spiral galaxies NGC~7232 and NGC~7233. This corresponds to \HI\ masses of (6.40 $\pm$ 0.35) $\times 10^{8}$ M$_{\odot}$ and (4.51 $\pm$ 0.46) $\times 10^{8}$ M$_{\odot}$, respectively. That makes a total flux of \FHI\ = (17.48 $\pm$ 0.51) Jy~km\,s$^{-1}$ (\HI\ mass of (2.31 $\pm$ 0.06) $\times 10^{9}$ M$_{\odot}$) for the galaxies in the triplet and \FHI\ = (33.53 $\pm$ 0.19) Jy~km\,s$^{-1}$ (\HI\ mass of (4.54 $\pm$ 0.03) $\times 10^{9}$ M$_{\odot}$) for the tidal H\,{\sc i} debris detected with MeerKAT. The newly discovered H\,{\sc i} debris is located around the two H\,{\sc i} clouds C1 and C2 as identified by \cite{2019MNRAS.487.5248L} with each cloud having a total flux of \FHI\ = (1.60 $\pm$ 0.09) and (4.41 $\pm$ 0.05) $\pm$ Jy~km\,s$^{-1}$. This corresponds to \HI\ masses of (2.17 $\pm$ 0.01) $\times 10^{8}$ M$_{\odot}$ and (5.90 $\pm$ 0.06) $\times 10^{8}$ M$_{\odot}$, respectively. \cite{2019MNRAS.487.5248L} reports only 7 Jy \kms associated with the \HI\ debris (clumps C1--6). MeerKAT detects $\sim$ five times more low column density H\,{\sc i} emission in filaments and clouds than ASKAP. The HIPASS total fluxes for the NGC~7232/3 triplet and its neighboring gas clouds are F$_{\rm HI}$ $\sim$46.0 Jy~km\,s$^{-1}$ \citep{10.1111/j.1365-2966.2004.07710.x} and F$_{\rm HI}$ $\sim$34.6 $\pm$ 4.1 Jy~km\,s$^{-1}$  \citep{2004AJ....128...16K}. MeerKAT hence recovers all \ion{H}{i} detected in Parkes single-dish observations in HIPASS and presumably the total \HI\ content of the NGC~7232 galaxy group (see Figure.~\ref{fig:profile}). Table~\ref{triplet1} summarises the \HI\ properties of the galaxy triplet and its surroundings. Table~\ref{triplet1} summarises the \HI\ properties of the galaxy triplet and its surroundings.

\subsection{H\,{\sc i} kinematics of the NGC~7232/3 triplet}
\begin{figure*}
\centering
   \includegraphics[height=10cm]{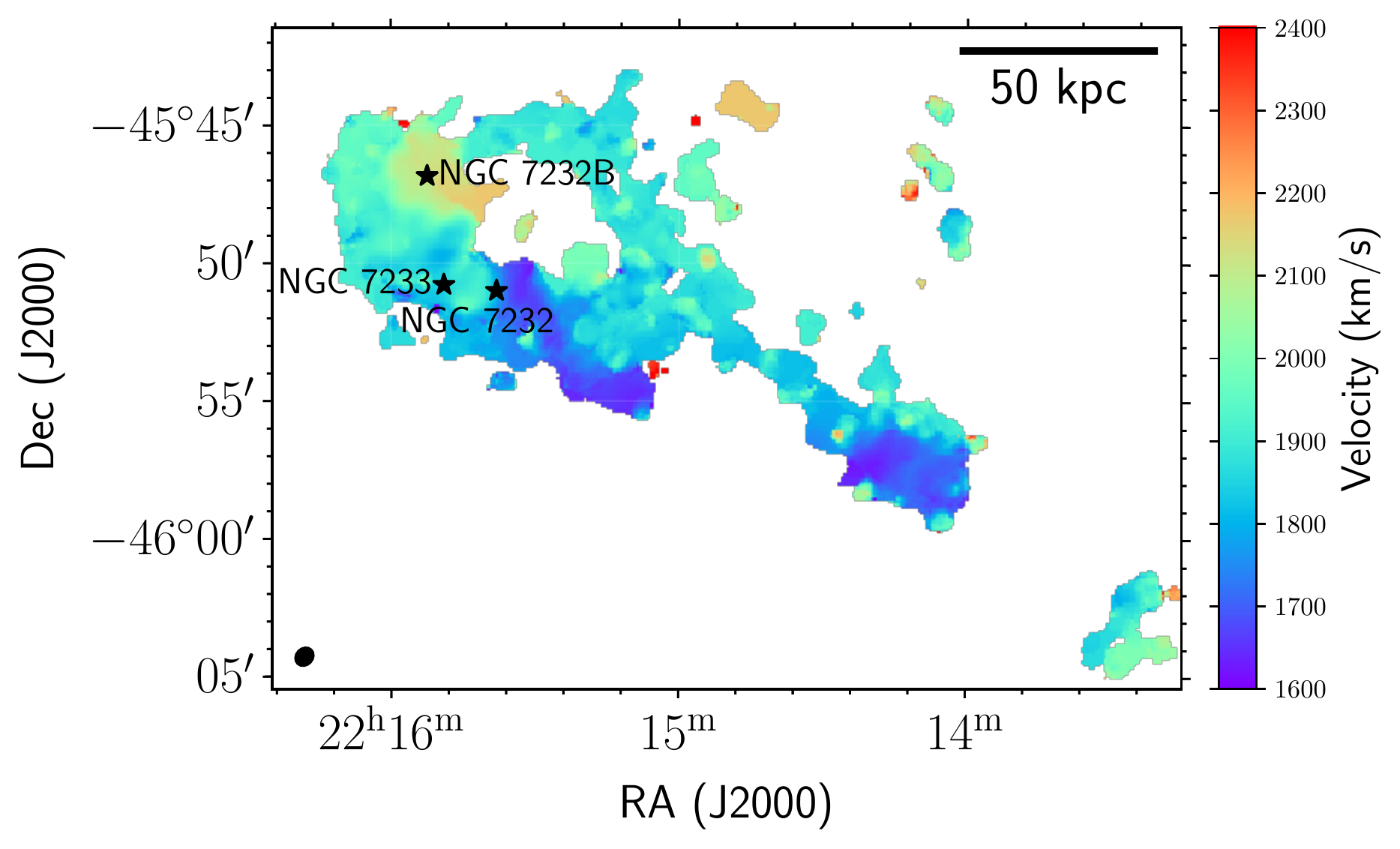}
\caption{MeerKAT-64 \HI\ velocity field map of the NGC~7232/3 triplet. The black stars indicate the optical centers of the triplet galaxies.}
\label{fig:triplet-mom}
\end{figure*}

Although the coarse velocity resolution of our observations does not allow for the detailed analysis of the kinematics of the triplet galaxy group and its surroundings, it is clear from the Fig.~\ref{fig:triplet-mom} that NGC~7232B has higher velocities and is kinematically different from NGC~7232 and NGC~7233 (velocity difference of up to 400 km\,s$^{-1}$). The \HI\ debris appears to have similar velocities to NGC~7233 and NGC~7232 within $\sim$200 km\,s$^{-1}$) which could indicate that most of this gas is coming from the interaction between the galaxy pair and not NGC~7232B.

\subsection{H\,{\sc i} deficiency equation of the NGC~7232/3 triplet}
We determine the \HI\ deficiency of each galaxy in the NGC~7232/3 triplet group to determine if the group spirals are \HI\ deficient or \HI\ rich. The \HI\ deficiency is derived by comparing the observed \HI\ mass to the expected \HI\ mass for an isolated galaxy. The expected H\,{\sc i} masses ($\text{M}_{\rm HI}^{\rm exp}$) are derived from the absolute B-magnitudes obtained from HyperLeda\footnote{http://leda.univ-lyon1.fr/} \citep{2014A&A...570A..13M}. We have used the scaling relation derived in \citep{2018A&A...609A..17J} based on measurements or constraints on the \HI\ masses of 844 isolated galaxies compiled by the AMIGA project (Analysis of the interstellar Medium in Isolated GAlaxies; \citealt{2005A&A...436..443V}).

\begin{equation}
\rm \log [M_{\text {H\,{\sc i}}}^{exp}/M_{\odot}] = 0.94 \log [L_{B}/L_{\odot}] + 0.18 \label{sec:equa}
\end{equation}
where $\log$ [L$_{\rm B}$/L$_{\odot}$] is derived by
\begin{equation}
\rm \log [L_{B}/L_{\odot}] = 10 + 2\log [D/\text{Mpc}] + 0.4(M_{\text{bol},\odot} - B_{c})
\end{equation}
where M$_{\text{bol},\odot}$ is the Sun\textquotesingle s bolometric absolute magnitude, we adopt M$_{\text{bol},\odot}$ = 4.88 \citep{2011A&A...534A.102L}, $D$ is the distance in Mpc and $B_{\rm c}$ is the absolute B-magnitude corrected for extinction. The  H\,{\sc i} deficiency is then defined by: 
\begin{equation}
\rm DEF = \log [M_{\rm HI}^{\rm exp}/M_{\odot}] - \log [M_{\rm HI}^{\rm obs}/M_{\odot}],
\end{equation}
where $\text{M}_{\rm HI}^{\rm obs}$ is the \HI\ mass derived from the observations. Positive values of DEF indicate \HI\ deficiency while negative values indicate \HI\ excess. We compare the expected \HI\ mass for each galaxy to its \HI\ mass  derived from our observations (see Table~\ref{triplet1}). NGC~7232 and NGC~7233 are \HI\ deficient while NGC~7232B is found to have excess \HI. Taking into account the uncertainties in $B$-magnitudes and the scatter in the scaling relation, we consider a galaxy to be \HI\ deficient when DEF $> 0.2$ and \HI\ excess if DEF $< -0.2$. The total \HI\ mass of the NGC~7232/3 triplet including all the \HI\ debris (i.e., \HI\ streams and clouds) is $\log [\frac{M_\ion{H}{i}}{ M_{\odot}}$] = 9.84 $\pm$ 0.002. Comparing the total derived \HI\ mass to the expected \HI\ mass of $\log [\frac{M_\ion{H}{i}}{ M_{\odot}}] = 9.61 \pm 0.11$, we find a total DEF of $-0.23 \pm 0.11$ which indicates that the NGC~7232/3 galaxy triplet has an overall \HI\ excess  compared to field galaxies.

\begin{table*} 
\centering
\caption{\HI\ properties of the NGC~7232/3 galaxy triplet and the \HI\ debris}
\label{tab}
\begin{tabular}{lccccc}
\hline
Galaxy & $F_{\rm HI}$ & $v_{\rm HI}$ & $\log$ [M$_{\rm HI}$ obs/M$_{\odot}$] &  $\log$ [M$_{\rm HI}$ pred/M$_{\odot}$] & \HI\ deficiency \\
 & (Jy~km\,s$^{-1}$)& (km\,s$^{-1}$) &  &  & \\ 
\hline
\hline        
NGC 7232B & 9.45 $\pm$ 0.31 & 2142 $\pm$ 17 & 9.11 $\pm$ 0.01 & 8.86 $\pm$ 0.21  & -0.25 $\pm$ 0.21 \\
NGC 7232  & 4.71$\pm$ 0.26 & 1740 $\pm$ 120 & 8.80 $\pm$ 0.01 & 9.31 $\pm$ 0.21 & 0.51 $\pm$ 0.21 \\
NGC 7233  & 3.32 $\pm$ 0.34 & 1958 $\pm$ 28 & 8.65 $\pm$ 0.03 & 9.12 $\pm$ 0.21 & 0.47 $\pm$ 0.21 \\
H\,{\sc i} debris & 33.53 $\pm$ 0.19 & 1963 $\pm$ 82& 9.65 $\pm$ 0.003 & \\
Triplet + debris& 51.01 $\pm$ 0.29 &1885 $\pm$ 115 &9.84 $\pm$ 0.002 &9.61 $\pm$ 0.11&-0.23 $\pm$ 0.11\\
\hline
	\multicolumn{6}{@{} p{15 cm} @{}}{\footnotesize{\textbf{Notes.} (1) Galaxy name, (2) integrated \HI\ flux, (3) \HI\ systemic velocity, (4) observed \HI\ mass, (5) predicted \HI\ mass derived using Equation.~\ref{sec:equa} and 6) calculated \HI\ deficiencies.}}
	\end{tabular}
    \label{triplet1}
\end{table*}

\begin{figure*}
\centering
   \includegraphics[width=15cm]{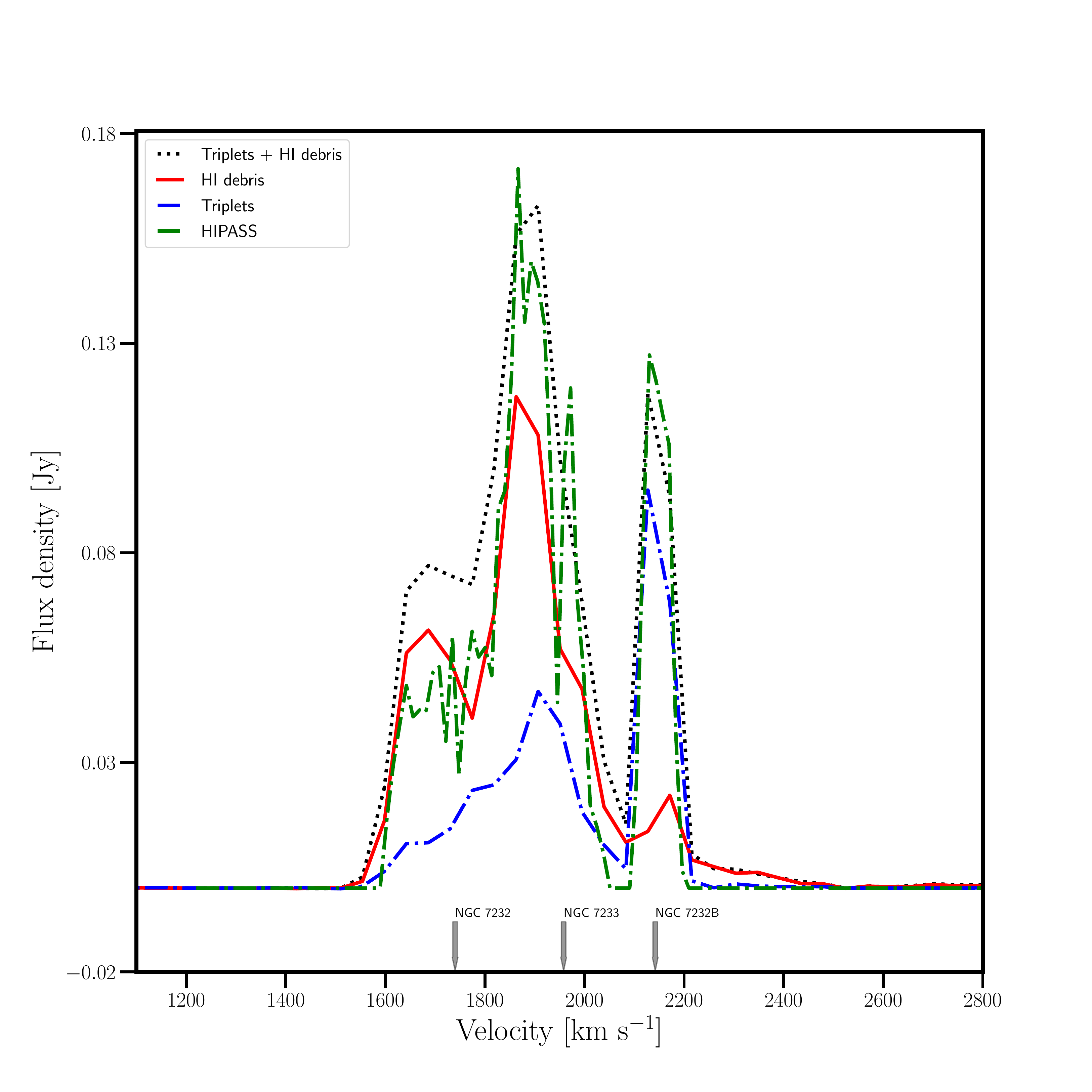}
\caption{MeerKAT-64 H\,{\sc i} global profile of both the NGC~7232/3 triplet and surrounding H\,{\sc i} debris (black dotted line), the H\,{\sc i} debris alone (red solid line), the galaxy triplet (blue dashed-dotted line) and the HIPASS global profile (green dashed-dotted line). The HIPASS global profile was extracted after applying a 4$\sigma$ threshold to the cube.}
\label{fig:profile}
\end{figure*}

\section{Discussion}\label{sec:discussion}
One of the questions coming from our data is the possible origin of the \HI\ debris observed. In this work we search for signs of interaction within the NGC~7323/3 galaxy triplet and its surrounding environment. The detection of wide-spread H\,{\sc i} debris around the galaxy triplet demonstrates the need for more sensitive, high resolution \HI\ observations if we are to have a complete understanding of how the low column density universe on small spatial scales contributes to the overall evolution of group galaxies.

\subsection{\HI\ content of the NGC~7232/3 triplet}
The presence of a large amount of H\,{\sc i} around the galaxy triplet, disturbed outer H\,{\sc i} discs, and the presence of a bridge connecting the three triplet members are clear evidence, showing that they are currently interacting within the last Gyr \citep{2001MNRAS.324..859B, 2004AJ....128...16K, 2019MNRAS.487.5248L}. One way of investigating whether galaxies in a group are undergoing transformation is by looking at their relative \HI\ content. We determine the \HI\ deficiency (DEF) of each galaxy. A comparison of the predicted H\,{\sc i} masses to the observed H\,{\sc i} masses is shown in Table~\ref{triplet1}. We find that NGC~7323 and NGC~7233 display an \HI\ deficiency while NGC~7232B is in excess. The \HI\ velocity field map (see Fig.~\ref{fig:triplet-mom} bottom) shows that NGC~7232B is significantly offset compared to NGC~7232 and NGC~7233. This could suggest that NGC~7232B has not lost a large amount of gas to the intergalactic medium of the triplet, and that most of the extended \HI\ gas probably originated from the galaxy pair. Considering the total \HI\ mass of the galaxy triplet and its surrounding, we find that the entire triplet is not deficient in \HI\ (-0.23 $\pm$ 0.11). \HI\ deficiencies or excess is not unique to group galaxies (see e.g.,  \citealt{2001A&A...377..812V,2005MNRAS.356...77K,2019A&A...632A..78J}). \cite{2001A&A...377..812V} studied a large sample of compact groups and proposed an evolutionary sequence that such groups follow. In the 1st phase the galaxies appear mostly undisturbed in HI, in the 2nd phase significant quantities of \HI\ appear in the intragroup medium, and in the final stage the majority of the \HI\ gas is either completely absent or outside the galaxies. Although that study did not consider triplets, the state of the NGC 7232/3 triplet is analogous to a phase 2 compact group. A similar result is seen for HGC~16 \citep{2019A&A...632A..78J}.

\subsection{Origin of the H\,{\sc i} debris around the NGC~7232/3 galaxy triplet}
The unsettled gas around the galaxy triplet strongly indicates interaction between the galaxies and their surrounding environment. The morphology of the newly detected H\,{\sc i} debris is complex and extends out to $\sim$ 140 kpc in projection. One prominent feature is an H\,{\sc i} tail which extends out to $\sim$ 94 kpc in projection. Given the length  of the \HI\ tail and the mean velocity dispersion ($\sigma$) of the triplet galaxy ($\sim$35 km\,s$^{-1}$), the age of the \HI\ tail must be at least 94 kpc/35 km\,s$^{-1}$ $\simeq$ 2.6 Gyr. The \HI\ tail and other surrounding \HI\ gas were not detected in the ASKAP observations by \citet{2019MNRAS.487.5248L} (see Figure ~\ref{fig:triplet-mom0}). The total H\,{\sc i} mass contained in the debris ($\sim$ 4.5 $\times$ 10$^{9}$ M$_{\odot}$) is more than two times the total H\,{\sc i} content of the galaxy triplet (see Figure ~\ref{fig:profile}).

Hydrodynamical simulations have shown that the group's tidal field can be responsible for the stripping of the \HI\ from the outer disc of gas-rich galaxies \citep{2005MNRAS.357L..21B}. The disturbed stellar- and gas morphologies associated with the NGC~7232/3 galaxy triplet is consistent with the scenario that the dominant mechanism in this group is via tidal stripping through galaxy-galaxy interactions. Such tails of \HI\ debris have been observed in numerous galaxy systems \citep{10.1093/mnras/sts033, 10.1093/mnras/stx2475, 2019A&A...632A..78J}.
 
The entire environment around the galaxy triplet including gas and stars (see Figure ~\ref{fig:triplet-mom}) does not cover a large volume. This could suggest that the system is gravitationally bound to the triplet. The cloud at the south of LEDA 130765 coincides in velocity quite well with the disc of the galaxy. The faded bridge/tail directed towards the triplet could indicate that this galaxy was involved in an earlier interaction with the triplet. Figure ~\ref{fig:profile} shows the global spectral profile of the galaxy triplet with its surrounding environment.

The profile indicates that most of the \HI\ in the bridge/tail is associated with the NGC~7332/3 galaxy pair and not NGC~7232B. One can also see from the profile that although the \HI\ gas is spatially disturbed it is well aligned in the spectral domain. While NGC~7232B appear to be more in the background based on their velocities (see Table~\ref{triplet1}), the velocities of the \HI\ features are consistent with that of NGC~7233 and NGC~7232. Thus, it is possible that the \HI\ debris are originating from NGC~7232 and NGC~7233 only. 
 
\begin{figure*}
\centering
   \includegraphics[width=5cm]{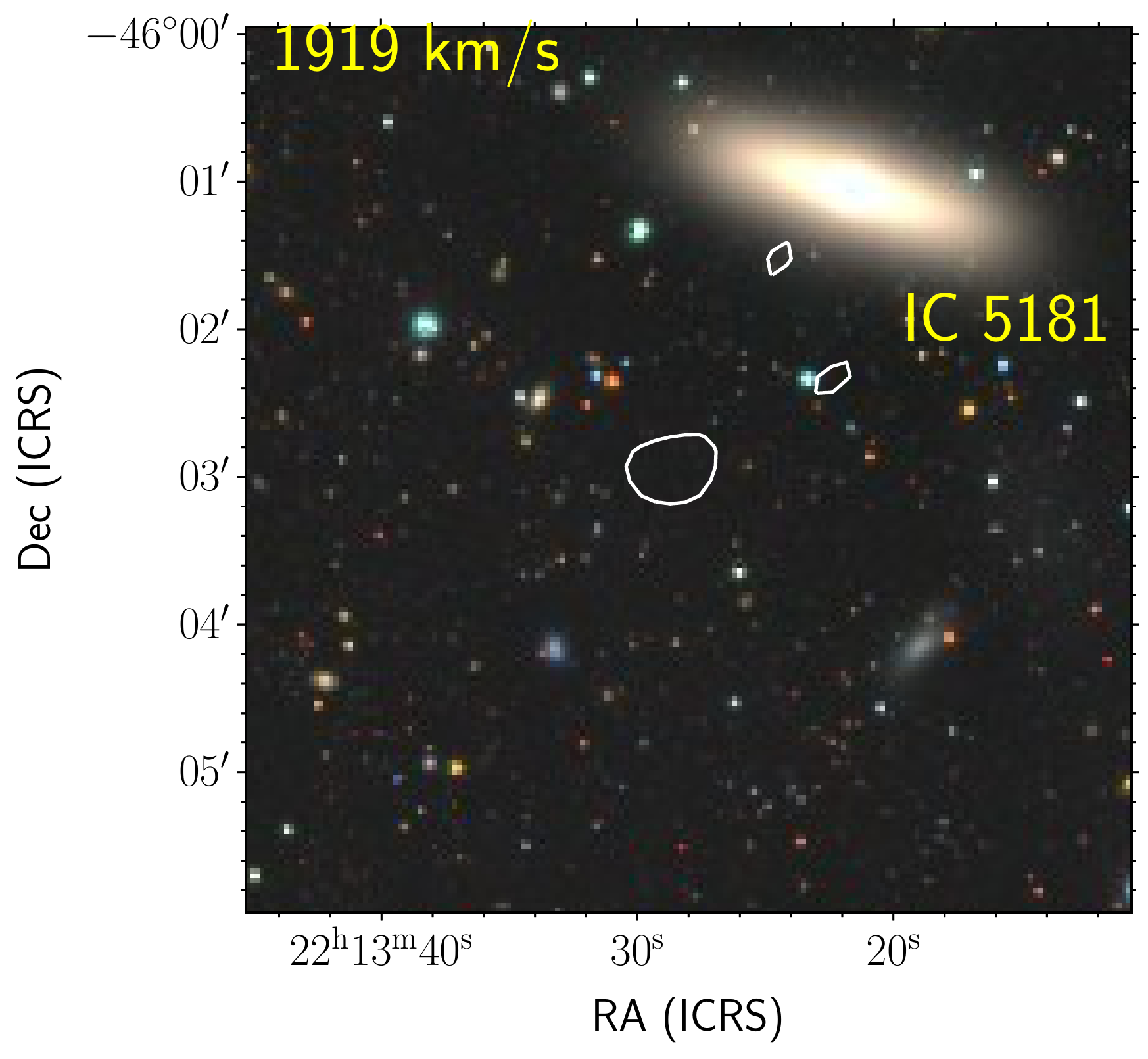}
   \includegraphics[width=5cm]{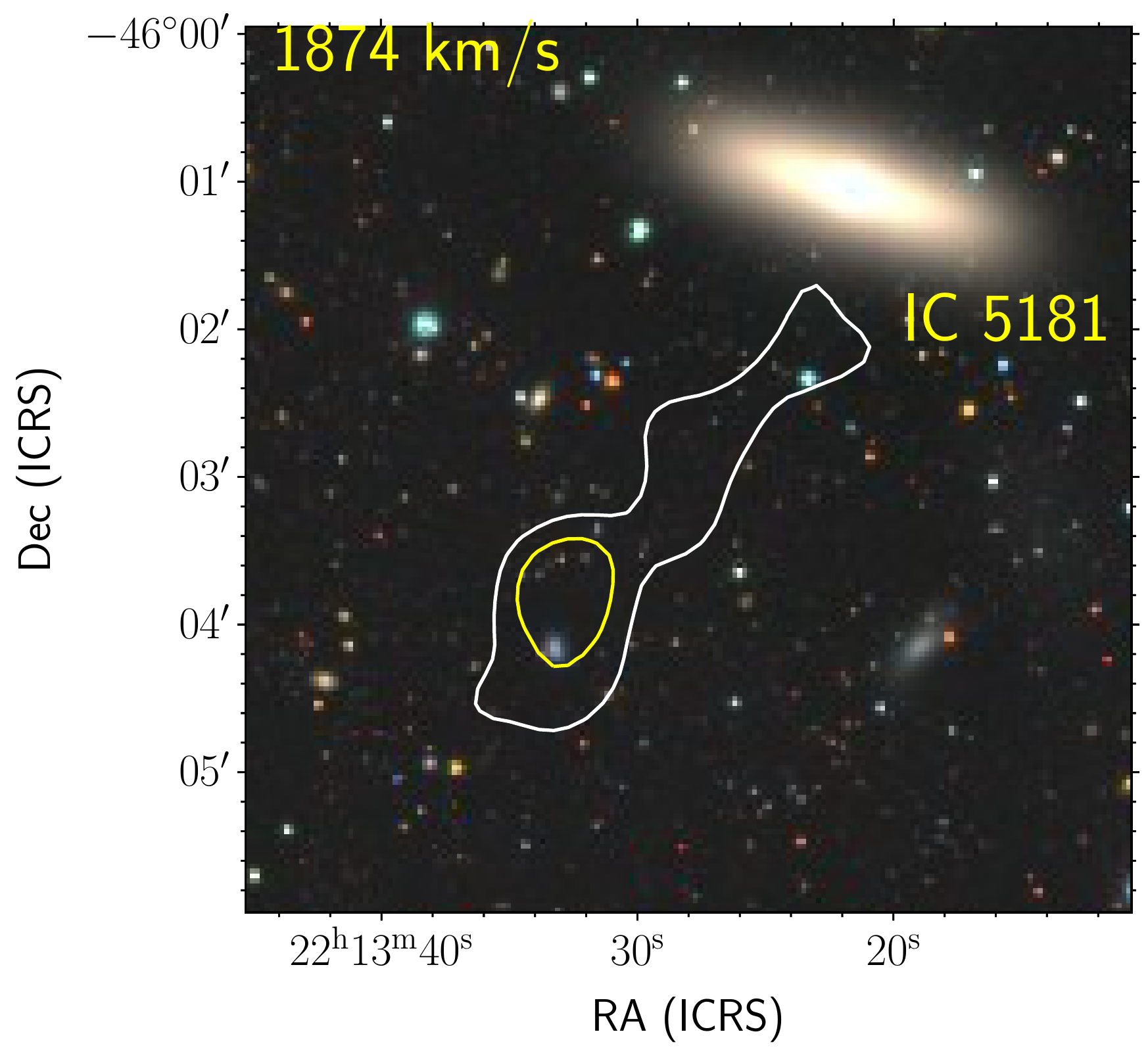}
   \includegraphics[width=5cm]{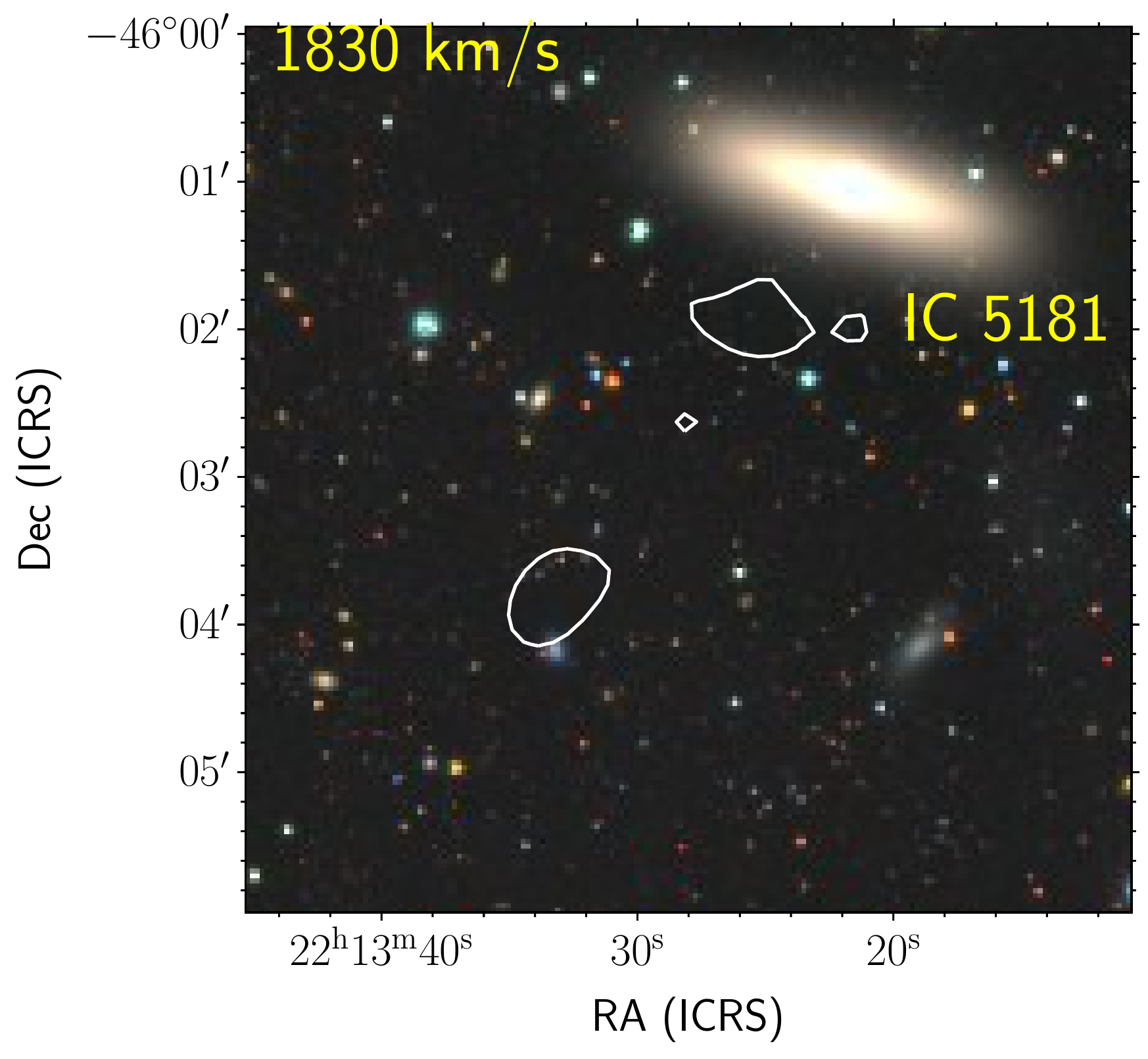}
\caption{MeerKAT-64 channel maps (primary beam corrected) showing the H\,{\sc i} cloud near the gas poor lenticular galaxy IC\,5181. The channel maps are overlaid on the DECaLS $grz$-band image. The contour colours are in increasing order with: white (1 $\times$ 10$^{19}$ cm$^{-2}$) and yellow (2.6 $\times$ 10$^{19}$ cm$^{-2}$), where (1 $\times$ 10$^{19}$ cm$^{-2}$) represent a 4$\sigma$ detection limit in the MeerKAT-64 \HI\ cube. The channel velocity is shown in the top left corner.}
\label{fig:cloud}
\end{figure*}

Figure ~\ref{fig:cloud} shows channel maps of the \HI\ cloud near IC\,5181. The \HI\ tail towards IC\,5181 is seen around 1874 km\,s$^{-1}$. The alignment of the lenticular galaxy with the \HI\ tail suggests that this galaxy could have interacted with the triplet and contributed some of its gas hence speeding up its own evolution. 

In short, our observations indicate that there is interaction taking place within the NGC~7232/3 galaxy triplet. However, in the future it will be interesting to combine \HI\ observations with simulations to determine the origin of the \HI\ debris seen with MeerKAT.

\subsection{H\,{\sc i} clouds around the NGC~7232/3 triplet}

Tidal interactions between gas-rich galaxies can create prominent gaseous tidal features which, in some cases, may form dense clumps of gas and possibly stars. Most of this tidally stripped material will eventually fall back into the interacting galaxies. However, during this process self-gravitating bodies, with comparably higher masses, can be formed out of the tidal debris. \citet{2006A&A...456..481B} show that as large tidally-formed clouds of \HI\ evolve, if they acquire sufficient \HI\ mass ( > 10$^{8}$ M$_{\odot}$) and move a sufficient distance away from their parent galaxies, they can decouple from their host tails and become kinematically distinct, long-lived tidal dwarf galaxies (TDGs). However, without the linking tidal tails, it becomes more difficult to discern a TDG from an accreted satellite dwarf.

We confirm the six \HI\ clumps detected with ASKAP \citep{2019MNRAS.487.5248L} and report one new faint \HI\ cloud located near the gas poor lenticular galaxy IC\,5181 (see Figure ~\ref{fig:triplet-mom0}). C3 and C4 are relatively low mass \HI\ clumps. We measure \HI\ masses of 6.7 $\times$ 10$^{7}$ M$_{\odot}$ and 9.5 $\times$ 10$^{7}$ M$_{\odot}$ for C3 and C4, respectively. 
The low \HI\ masses of C3 and C4 show that these \HI\ clumps are likely short-lived and will probably be reaccreted by the parent spiral galaxies. These two \HI\ clumps are located within the low column density complex of \HI\ emission surrounding the galaxy triplet, which makes it difficult to determine their exact origin.

C2 is the most gas-rich \HI\ clump detected near the galaxy triplet with an \HI\ mass of 5.9 $\times$ 10$^{8}$ M$_{\odot}$. C2 appears to be embedded in the same \HI\ envelope as C1 (see Figure ~\ref{fig:triplet-mom0}), which has an \HI\ mass of 2.2 $\times$ 10$^{8}$ M$_{\odot}$. The two clumps are found within an \HI\ tail that is oriented parallel to the gas-poor lenticular galaxy IC\,5181. C1 is just reaching the minimum \HI\ mass threshold that is needed to form a long-lived TDG \citep{2006A&A...456..481B}, meaning that it is likely a transient tidal feature. On the other hand, the \HI\ mass of C2 is well above 10$^{8}$ M$_{\odot}$. This implies that C2 has the potential to develop into the self-gravitating long-lived TDG. In addition to its \HI\ mass, C2 appears to be at the tip of a tidal tail and is spatially distanced from its assumed parent galaxies. The lack of optical counterparts associated with C2 could indicate that the system is still young with respect to the on-going interaction event.

\begin{figure*}
\centering
   \includegraphics[width=8cm]{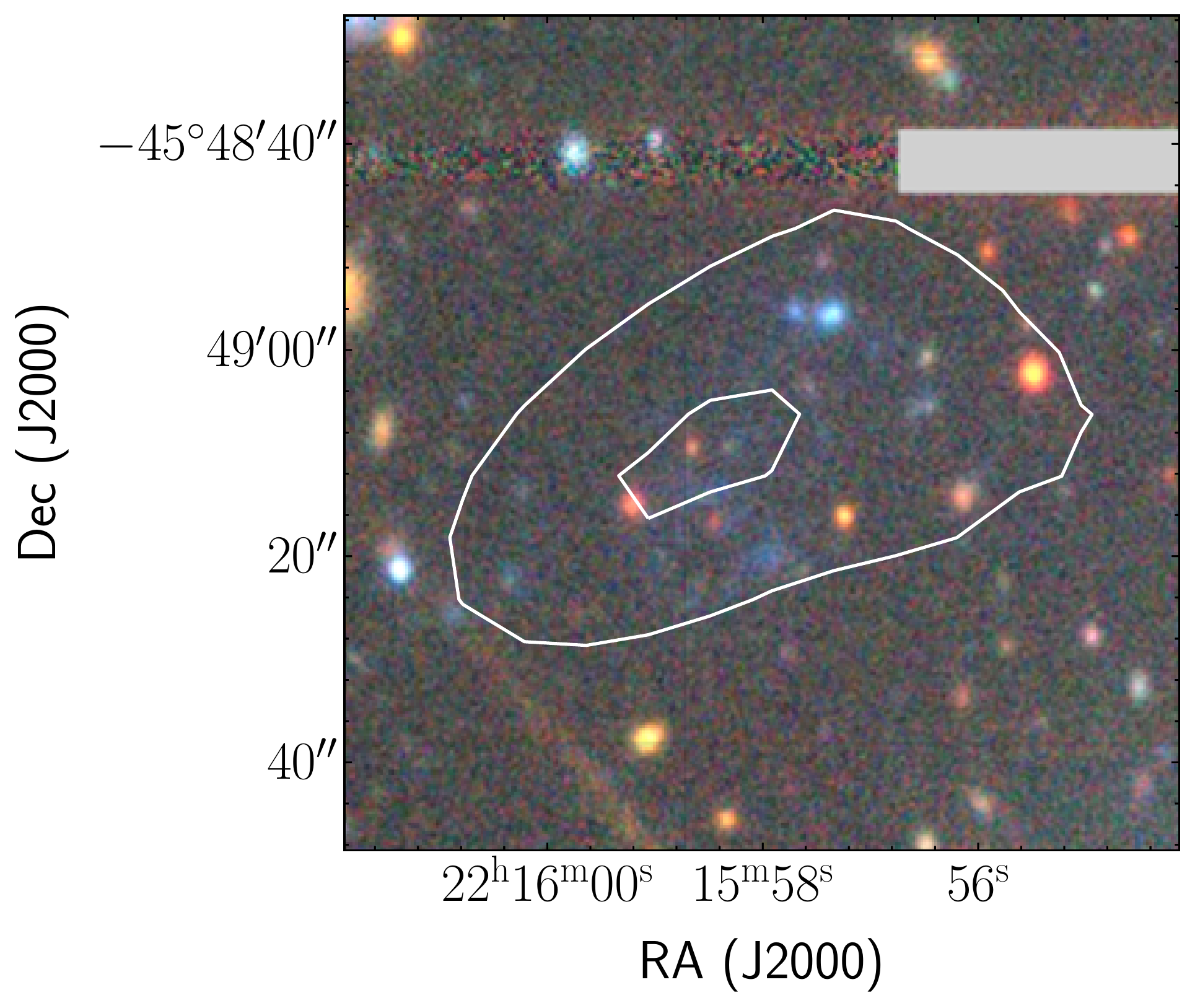}
   \includegraphics[width=8cm]{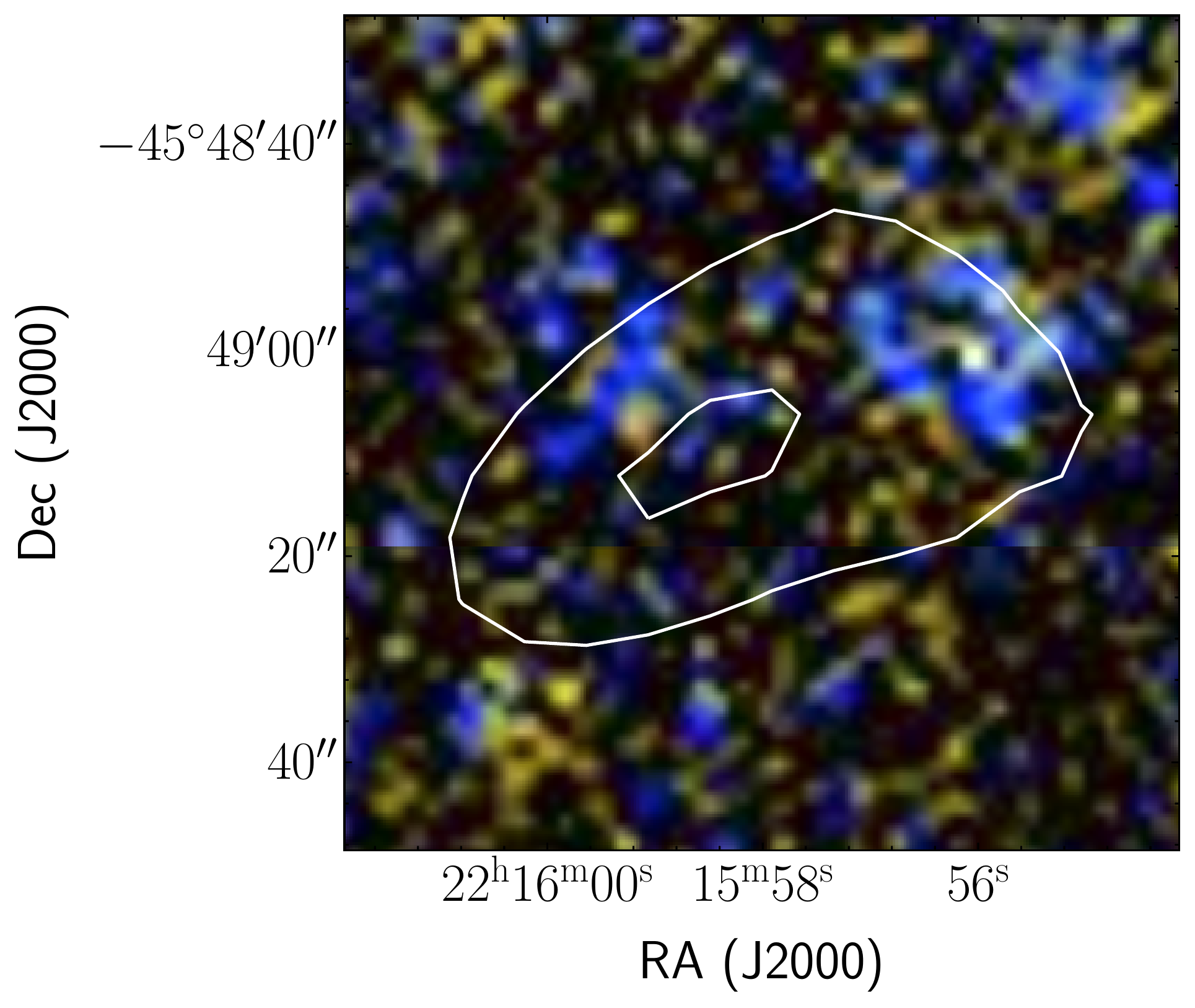}
\caption{{\bf Left panel:} MeerKAT-64 integrated H\,{\sc i} intensity map of the C6 cloud overlaid on the DECaLS $grz$-band image. Faint blue features within the \HI\ contour could indicate potential in-situ star formation within the \HI\ cloud C6. {\bf Right panel:} MeerKAT-64 integrated \HI\ intensity map of the C6 cloud overlaid onto a combined GALEX NUV/FUV image. The blue colour (FUV) indicates the presence of young stellar populations within the \HI\ contour. The contours are (5.0, 6.5) $\times$ 10$^{20}$ cm$^{-2}$.}
\label{fig:cloud1}
\end{figure*}

The \HI\ cloud C5 is located close to the spiral galaxy NGC~7232B and has an \HI\ mass of 2.2 $\times$ 10$^{8}$ M$_{\odot}$. Its velocity is $\sim$1876 km$^{-1}$ similar to the \HI\ gas associated with NGC~7232 and NGC~7233. Its position suggests that it is a result of an interaction between the spiral galaxies NGC~7232B and NGC~7233. The location of C5 and its \HI\ mass (just slightly above 10$^{8}$ M$_{\odot}$) suggests that this cloud is likely to fall back into the parent galaxies. The \HI\ cloud C6 is located between NGC~7233 and NGC~7232B and is aligned opposite to C5. C6 has an \HI\ mass of 4.2 $\times$ 10$^{8}$ M$_{\odot}$ and its peak column density is $\sim$ 10$^{20}$ cm$^{-2}$. The proximity of C6 to its parent galaxies implies that the \HI\ cloud is likely to fall back into the parent galaxies \citep{2006A&A...456..481B}. However, as the interaction between NGC 7232B and NGC 7233 is still actively ongoing, it is also possible that the tail containing C6 might extend further, moving C6 outwards. If this happens and C6 retains or even gains more \HI\ mass, there is a possibility that it might evolve into a long-lived TDG. As indicated in \citet{2007Sci...316.1166B}, not all potential TDGs are located at the tip of the tails so the (projected) position of C6 within the tidal tail might be less of a factor in its development. Optical and UV images of show faint blue features within the peak \HI\ column density contours of C6 (see Figure ~\ref{fig:cloud1} left panel). These optical features are more pronounced in UV (see Figure ~\ref{fig:cloud1} right panel), thus suggesting the presence of newly formed stars.

\section{Summary}\label{sec:summary}
We present results from the MeerKAT open time science observations obtained in L-band (20-cm) using the 4k-wideband mode (44 km \,s$^{-1}$ velocity resolution). The aim of the observations was to search for \HI\ emission associated with the NGC~7232/3 galaxy triplet and provide an understanding to the interaction processes between the triplet with its surrounding environment. With a 4$\sigma$ column density sensitivity of 1 $\times$ 10$^{19}$ cm$^{-2}$, we find extended \HI\ debris around the NGC~7232/3 galaxy triplet. The newly discovered H\,{\sc i} streams extend over $\sim$20 arcmin (140~kpc) in projection, and contain an H\,{\sc i} mass of $\sim$4.5 $\times 10^9$~M$_{\odot}$, more than 50\% of the total  H\,{\sc i} mass of the galaxy triplet. This result provide a new understanding of where the gas resides within the galaxy triplet and clearly shows how galaxy interactions play an important role in the evolution of galaxies in the NGC~7232/3 galaxy triplet.

The presence of low column density \HI\ as well as the distorted morphologies of the triplet galaxies to some extent indicate galaxy interactions. We find that NGC~7232 and NGC~7233 have lost the vast majority of their expected \HI\ mass as opposed to NGC~7232B which seems to have retained its \HI. The velocities of each galaxy in the triplet show that NGC~7232B is significantly offset as compared to NGC~7233 and NGC~7232 and that the velocities of the \HI\ debris are consistent with the galaxy pair NGC~7232 and NGC~7233. This suggests that NGC~7232B contains excess \HI\ because it has not really lost a large amount of gas to the intragalactic medium of the triplet, and that most of the extended \HI\ debris gas originated from the galaxy pair. Overall, we find that the galaxy triplet is not \HI\ deficient. The state of the NGC 7232/3 triplet is analogous to a phase 2 compact group. A similar result has been found for HGC~16 \citep{2019A&A...632A..78J}.

We detected 15 extragalactic \HI\ sources in our data cube and confirmed the detection of 6 individual \HI\ clouds from previous ASKAP observations. We report one new detection , an \ion{H}{i} counterpart to the galaxy LEDA~130765 (WISEA J221403.60--454843.7) which is possibly a dwarf transition galaxy with an \HI\ mass of 4.8 $\pm$ 1.0 $\times$ 10$^{7}$ M$_{\odot}$. 

Based on the success of our MeerKAT-64 observations, we look forward to a much larger project mapping the star formation and H\,{\sc i} gas distribution within nearby galaxy groups to assess their evolutionary state and formation history. The availability of high velocity resolution in future will allow for much advanced kinematic analysis of galaxy dynamics and multi-wavelength studies to investigate the interaction processes occurring in the NGC~7232/3 galaxy group.

\section{Data availability}
The data from this study are available upon request to the corresponding author, Brenda Namumba.

\section*{Acknowledgements}
The MeerKAT telescope is operated by the South African Radio Astronomy Observatory, which is a facility of the National Research Foundation, an agency of the Department of Science and Innovation. BN's research is supported by the South African Radio Astronomy Observatory (SARAO). We acknowledge the Inter-University Institute for Data Intensive Astronomy (IDIA) for supporting us with the data intensive cloud for data processing. IDIA is a South African university partnership involving the University of Cape Town, the University of Pretoria and the University of the Western Cape. BN acknowledges financial support from the CSIC Program of Scientific Cooperation for Development i-COOP+2019. The research of OS is supported by the South African Research Chairs Initiative of the Department of Science and Technology and National Research Foundation. PK is partially supported by the BMBF project 05A17PC2 for D-MeerKAT. LVM, MGJ and BN acknowledge financial support from the State Agency for Research of the Spanish MCIU through the "Center of Excellence Severo Ochoa" award to the Instituto de Astrofísica de Andalucía (SEV-2017-0709). BN acknowledge the discussion with the AMIGA team at IAA regarding data reduction and analysis. LVM, MJ, SSE and JG acknowledge as well support from the grants AYA2015-65973-C3-1-R (MINECO/ FEDER, UE) and RTI2018-096228-B-C31(MCIU/AEI/FEDER,UE). MGJ fellowship was supported by a Juan de la Cierva formaci\'on fellowship. The work of WJGdB has received funding from the European Research Council (ERC) under the European Union's Horizon 2020 research and innovation programme (grant agreement No 882793/MeerGas). This project has received funding from the European Research Council (ERC) under the European Union’s Horizon 2020 research and innovation programme (grant agreement no. 679627; project name FORNAX). MEC is a recipient of an Australian Research Council Future Fellowship (project No. FT170100273) funded by the Australian Government.



\section*{Appendix}
\subsection*{\HI\ detected galaxies in NGC~7232 field}\label{sec:meerkatdetections}
\begin{figure*}
\centering
   \includegraphics[width=13cm]{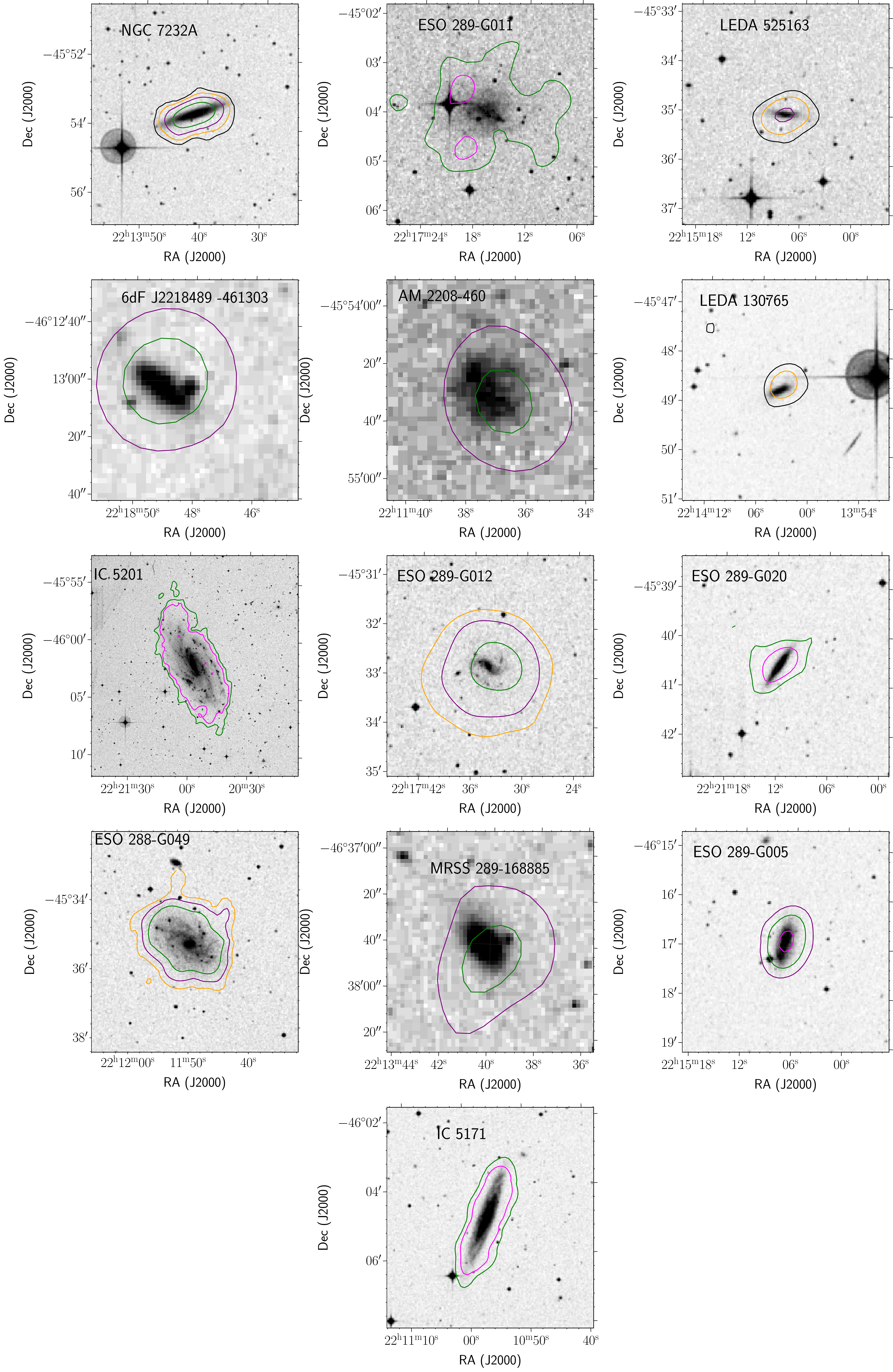}
\caption{MeerKAT integrated \HI\ column density maps (primary beam corrected) of individual galaxies in the NGC~7232 group overlaid on DSS optical images. The contour levels are 1 $\times$ 10$^{19}$ cm$^{-2}$ $\times$ (2, 4, 8, 16, 32), where 1 $\times$ 10$^{19}$ cm$^{-2}$ represent the 4$\sigma$ detection limit. The colours are in order of increasing column density: black, orange, purple, green and magenta.}
\label{fig:1}
\end{figure*}

\begin{figure*}
\centering
   \includegraphics[width=13cm]{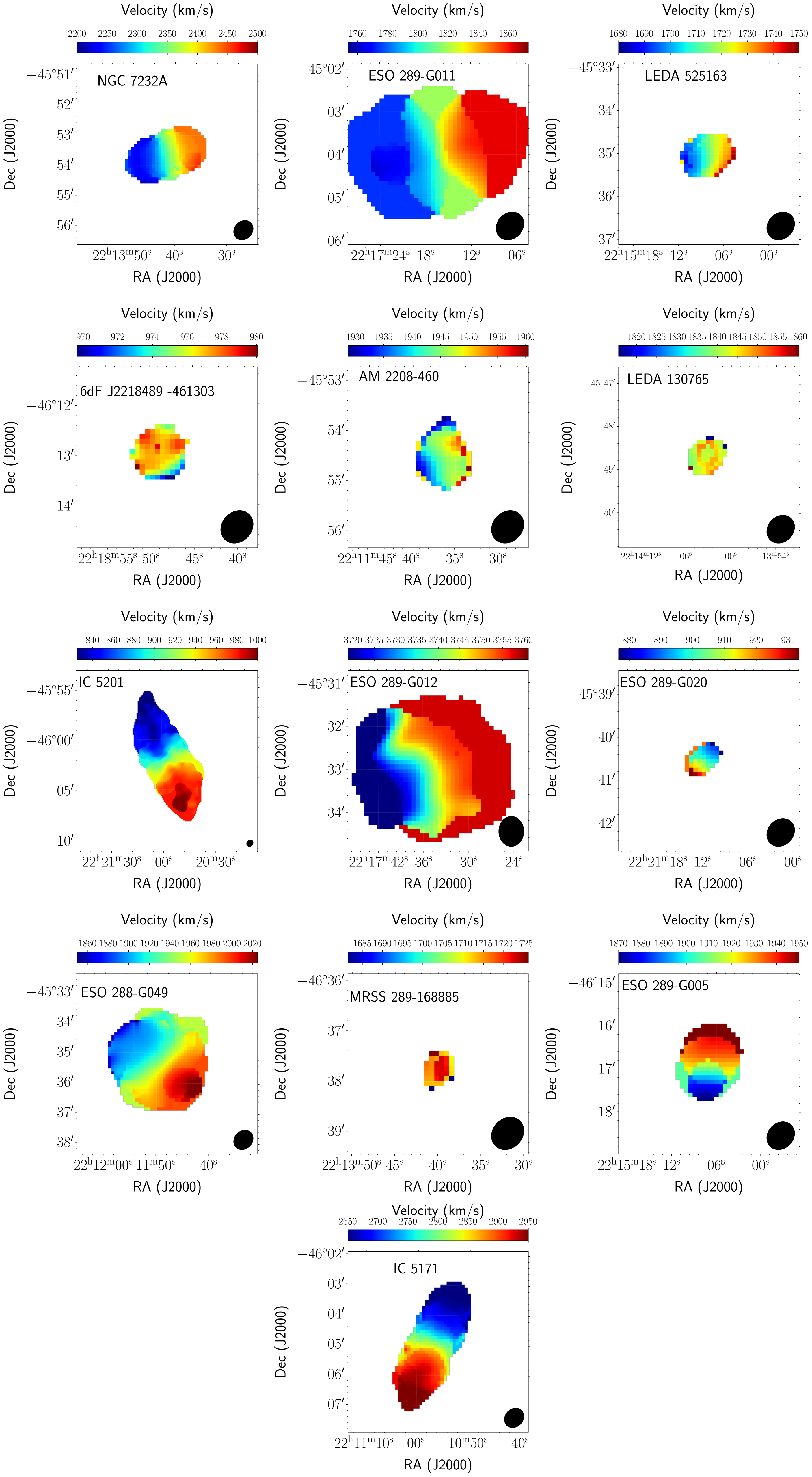}
\caption{MeerKAT mean \HI\ velocity fields of individual galaxies in the NGC~7232 galaxy group. The synthesized beam is shown in the bottom right corner.}
\label{fig:2}
\end{figure*}

\begin{figure*}
\centering
   \includegraphics[width=15cm]{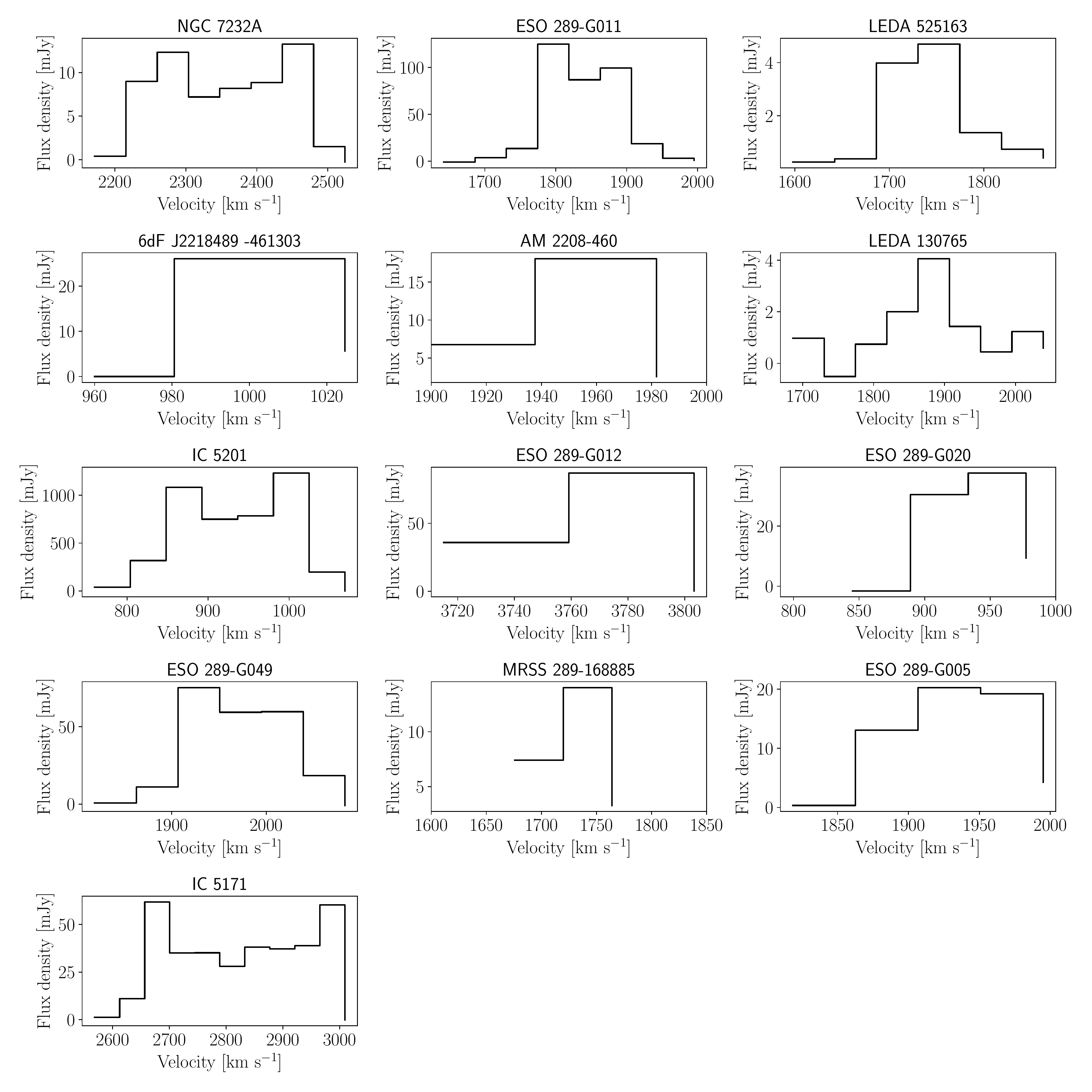}
\caption{MeerKAT \HI\ intensity profiles of individual galaxies in the NGC~7232 group obtained using SoFiA, i.e. only \HI\ signal deemed reliable is shown.}
\label{fig:3}
\end{figure*}

We detect 16 galaxies in \HI\ over a velocity range of $\sim$700 to 4000~km\,s$^{-1}$. For comparison, \cite{2019MNRAS.487.5248L} studied the H\,{\sc i} emission in the NGC~7232 galaxy group over a velocity range of $\sim$1500 to 3000~km\,s$^{-1}$, while \cite{10.1093/mnras/stz2063} focused on the foreground galaxy IC\,5201 and its surroundings (focusing on the velocity range from --20 to 2513~km\,s$^{-1}$), both using selected pointings from the first WALLABY Early Science observations. In this section we briefly describe the general properties of the  H\,{\sc i}-detected galaxies in our MeerKAT-64 observations. Table~\ref{all} summarizes the \HI\ properties of the galaxies and compares with ASKAP \citep{2019MNRAS.487.5248L} and HIPASS \cite{2004AJ....128...16K}. Figure ~\ref{fig:1}, ~\ref{fig:2}, and ~\ref{fig:3} show the integrated maps, velocity field maps and global profiles of each galaxy. 
\begin{itemize}
\item The nearby spiral galaxy IC\,5201 lies in the foreground of the NGC~7232 group, and was studied in detail by \cite{10.1093/mnras/stz2063} using WALLABY Early Science data. IC\,5201 is detected at the eastern edge of our MeerKAT-64 field. We derive an H\,{\sc i} mass of (7.90 $\pm$ 0.04) $\times$ 10$^{9}$ M$_{\odot}$ using an adopted distance of 13.2 Mpc \citep{10.1093/mnras/stz2063}. We measure a systemic velocity of (914 $\pm$ 52) km\,s$^{-1}$.

\item ESO 289-G020 is a companion galaxy of IC\,5201, also detected by \cite{10.1093/mnras/stz2063}. It is an edge-on galaxy with a systemic velocity of (925 $\pm$ 14) km\,s$^{-1}$. We measure an H\,{\sc i} mass of (1.4 $\pm$ 0.2) $\times$ 10$^{8}$ M$_{\odot}$ (assuming $D$ = 13.2 Mpc). Our derived \HI\ mass is in agreement with the ASKAP observations \citep{10.1093/mnras/stz2063}. 

\item For the spiral galaxy ESO 288-G049 we find a systemic velocity of ($\sim$1950 $\pm$ 15) km\,s$^{-1}$ and calculate an H\,{\sc i} mass of (1.3 $\pm$ 0.5) $\times$ 10$^{8}$ M$_{\odot}$.\cite{2019MNRAS.487.5248L} report an ASKAP \HI\ mass which is $\sim$ 39$\%$ less than our MeerKAT value.  

\item For the edge-on spiral galaxy IC\,5171 we find a systemic velocity of ($\sim$2801 $\pm$ 115) km\,s$^{-1}$ and an \HI\ mass of (5.4 $\pm$ 0.3) $\times$ 10$^{9}$ M$_{\odot}$ ($D$ = 38.8 Mpc). Following \citep{2019MNRAS.487.5248L}, MeerKAT detects $\sim$27\% more \HI\ mass than ASKAP. The HIPASS \HI\ mass is in good agreement with our MeerKAT results. 
 
\item For the irregular dwarf galaxy ESO 289-G011, we measure a systemic velocity of (1816 $\pm$ 36) km\,s$^{-1}$. Adopting a distance of 24 Mpc, we derive an H\,{\sc i} mass of (2.10 $\pm$ 0.04) $\times$ 10$^{9}$ M$_{\odot}$. MeerKAT detects more \HI\ mass when compared with ASKAP \citep{2019MNRAS.487.5248L}. 

\item The spiral galaxy ESO 289-G005 was first detected in WALLABY Early Science data by  \cite{2019MNRAS.487.5248L}. We measure a systemic velocity of (1913 $\pm$ 25) km\,s$^{-1}$. An \HI\ mass of (3.4 $\pm$ 0.4) $\times$ 10$^{8}$ M$_{\odot}$ is derived (D = 24 Mpc). Our derived \HI\ mass is in agreement with the ASKAP reported by \cite{2019MNRAS.487.5248L}. 
  
\item AM 2208--460 is a dwarf galaxy, first detected in H\,{\sc i} in WALLABY Early Science data \citep{2019MNRAS.487.5248L}. We measure a systemic velocity of ($\sim$1958 $\pm$ 7) km\,s$^{-1}$. No  velocity from optical spectroscopy exists in the literature. Adopting a distance of 24 Mpc, we derive an H\,{\sc i} mass of (1.8 $\pm$ 0.3) $\times$ 10$^{8}$ M$_{\odot}$.
Compared to the ASKAP, we derive $\sim$ 50\% more H\,{\sc i} mass.

\item LEDA~525163 is a dwarf irregular galaxy. Here, we measure a systemic velocity of (1715 $\pm$ 15) km\,s$^{-1}$. We derive an H\,{\sc i} mass of (6.1 $\pm$ 1.0) $\times$ 10$^{7}$ M$_{\odot}$ for a distance of 24 Mpc.

\item 6dF J2218489--461303 (PGC 130771) is a foreground dwarf irregular galaxy near IC\,5201. We find H\,{\sc i} emission centered around (984 $\pm$ 1) km\,s$^{-1}$. We derive an H\,{\sc i} mass of (1.5 $\pm$ 0.2) $\times$ 10$^{8}$ M$_{\odot}$ ($D$ = 24 Mpc). \cite{10.1093/mnras/stz2063} reports an \HI\ mass $\sim$44$\%$ less. 

\item MRSS 289-168885  is possibly a dwarf transition galaxy. Its optical appearance is spheroidal and undisturbed. We find H\,{\sc i} emission centered around (1721 $\pm$ 9) km\,s$^{-1}$. Adopting a distance of 24 Mpc, we derive an H\,{\sc i} mass of (8.4 $\pm$ 2.3) $\times$ 10$^{7}$ M$_{\odot}$. 

\item NGC~7232A is an edge-on spiral galaxy (type SA), not catalogued in HIPASS. We find H\,{\sc i} emission centered around (2320 $\pm$ 78( km\,s$^{-1}$. We derive an H\,{\sc i} mass of (3.5 $\pm$ 0.4) $\times$ 10$^{8}$ M$_{\odot}$ ($D$ = 24 Mpc), similar to the ASKAP \citep{2019MNRAS.487.5248L}. Optical images of NGC~7232A show it to be symmetric and undisturbed. 

\item LEDA~130765 (WISEA J221403.60--454843.7), not mentioned by \cite{2019MNRAS.487.5248L}, is possibly a dwarf transition galaxy located just north of the extended tidal debris seen in Fig.~\ref{fig:triplet-mom}. Its optical appearance is lenticular and undisturbed. We find H\,{\sc i} emission centered around (1824 $\pm$ 7) km\,s$^{-1}$ and measure an H\,{\sc i} mass of (4.8 $\pm$ 1.0) $\times$ 10$^{7}$ M$_{\odot}$ ($D$ = 24 Mpc).  

\item The galaxies LEDA 519373 and 519377 (around 2008 km\,s$^{-1}$) are located just south of the lenticular galaxy IC\,5181 near the debris at the south-western end of the extended H\,{\sc i} filament. We measure a total H\,{\sc i} flux of $F_{\rm HI} \sim 1.83 \pm 0.05$ Jy~km\,s$^{-1}$ in that region.

\item ESO 289-G012 is a spiral galaxy behind the NGC~7232 group. We measure a systemic velocity of (3564 $\pm$ 15) km\,s$^{-1}$ and calculate an H\,{\sc i} mass of (3.6 $\pm$ 0.1) $\times$ 10$^{9}$ M$_{\odot}$ ($D$ = 53 Mpc). 
\end{itemize}

\clearpage
\begin{sidewaystable}
\centering
\caption{Other \HI\ sources detected with MeerKAT in the NGC~7232 galaxy group.}\label{tab3}
\begin{adjustbox}{width=\textwidth}
\begin{tabular}{lccccccccccc}
\hline
Galaxy &Type &\multicolumn{2}{c}{$\alpha,\delta$ (J2000)} & Optical $v_{\rm hel}$& HIPASS $v_{\rm HI}$ & HIPASS F$_{\rm HI}$ & ASKAP $v_{\rm HI}$ & ASKAP F$_{\rm HI}$ & MeerKAT $v_{\rm HI}$ &MeerKAT F$_{\rm HI}$ & MeerKAT M$_{\rm HI}$\\
 & &(hms) &(dms) &(km\,s$^{-1}$) & (km\,s$^{-1}$) & (Jy~km\,s$^{-1}$) & (km\,s$^{-1}$) & (Jy~km\,s$^{-1}$) & (km\,s$^{-1}$) & (Jy~km\,s$^{-1}$) &(10$^9$ M$_\odot)$ \\   
\hline        
IC 5201		& SB(rs)cd& 22:20:57.44 & --46:02:09.1  & 914 & -- & -- &    914  & 177 $\pm$ 1 & 914 $\pm$ 15 & 194 $\pm$ 1.00 & 7.90 $\pm$ 0.04 \\ 
ESO 289-G020 & --& 22:21:11.71 &--45:40:35.5  & 911 &-- & --& --&  --  & 925 $\pm$ 14 & 3.60 $\pm$ 0.60 &0.14 $\pm$ 0.02  \\  
6dF J2218489--461303 &--&22:18:48.93 & --46:13:03.1 & 1207 &  -- &-- & 996 $\pm$ 2 &0.6 $\pm$ 1 & 984 $\pm$ 1 & 1.09 $\pm$ 0.21 &0.15 $\pm$ 0.02  \\
\hline
ESO 288-G049 &SB(rs)dm&22:11:48.56 & --45:35:22.7&1968 & 1964 & 6.6 & 1964 & 6 $\pm$ 1 & 1950 $\pm$ 15 &9.87 $\pm$ 0.40 &0.13 $\pm$ 0.05 \\
IC 5171	 &SAB(rs)bc&22:10:56.70 & --46:04:53.3&2847& 2836 & 15.8 & 2831 & 11 $\pm$ 2 & 2801 $\pm$ 115 & 15.27 $\pm$ 0.82 & 5.40 $\pm$ 0.30 \\ 

ESO 289-G011	&IB(s)m & 22:17:16.33& --45:03:59.4 &1818 & 1817 & 10.5 & 1821 & 8 $\pm$ 2 & 1816 $\pm$ 36& 15.4 $\pm$ 0.29& 2.10 $\pm$ 0.04 \\ 

ESO 289-G005	& --& 22:15:07.16 & --46:16:53.4 & 1885 & --& 11.9 & 1931 & 2.1 $\pm$ 0.4 & 1913 $\pm$ 25 & 2.52 $\pm$ 0.33 & 0.34 $\pm$ 0.04 \\ 

AM 2208--460	& dwarf &22:11:37.13 & --45:54:31.0 &  --	& -- &-- &  1953&0.6 $\pm$ 0.1 &1958 $\pm$ 7&1.19 $\pm$ 0.20 &0.18 $\pm$ 0.03 \\ 

LEDA 525163	 &-- &22:15:08.01 & --45:35:01.4 & --&-- &-- &--&-- &1715 $\pm$ 15 &0.45  $\pm$ 0.11 &0.06 $\pm$ 0.01 \\

MRSS 289-168885    & --& 22:13:40.24& --46:37:39.9 &-- & --&--& 1741&0.5 $\pm$ 0.1 &1721 $\pm$ 9 &0.62 $\pm$ 0.17 &0.08 $\pm$ 0.02 \\ 

NGC 7232A   & SB(rs)ab&22:13:41.44& --45:53:37.4 &2407& --& -- &2348 &2.0 $\pm$ 0.4 &2320 $\pm$ 78& 2.56 $\pm$ 0.32&0.35 $\pm$ 0.043 \\  

LEDA 130765 &-- & 22:14:03.63 & --45:48:43.6 &-- &-- &-- &--& -- &1824 $\pm$ 7 &0.35 $\pm$ 0.07& 0.05 $\pm$ 0.01\\ 
\hline
ESO 289-G012 & dwarf spiral & 22:17:34.29 & --45:32:48.8 &3792  & -- &--&-- & -- & 3564 $\pm$ 15 & 5.46 $\pm$ 0.21 & 3.60 $\pm$ 0.14 \\
\hline 
\end{tabular}
  \end{adjustbox}
\label{all} 
\\

{{\bf Notes.} (1) Galaxy name, (2) Galaxy type \cite{1991rc3..book.....D}, (3) RA and DEC \cite{1991rc3..book.....D} and (4) optical $v_{\rm hel}$
\cite{1991rc3..book.....D}, Cols. (5) \& (6:) \citet{10.1111/j.1365-2966.2004.07710.x}, Cols. (7) \& (8): \citet{2019MNRAS.487.5248L}, Cols. (9) -- (11): this work.\\ For IC 5201 and 6dF J2218489--461303 Cols. (7) \& (8): \citet{10.1093/mnras/stz2063}}
\end{sidewaystable}
\clearpage

\input{final.bbl}




\begin{thebibliography}{}
\makeatletter
\relax
\def\mn@urlcharsother{\let\do\@makeother \do\$\do\&\do\#\do\^\do\_\do\%\do\~}
\def\mn@doi{\begingroup\mn@urlcharsother \@ifnextchar [ {\mn@doi@}
  {\mn@doi@[]}}
\def\mn@doi@[#1]#2{\def\@tempa{#1}\ifx\@tempa\@empty \href
  {http://dx.doi.org/#2} {doi:#2}\else \href {http://dx.doi.org/#2} {#1}\fi
  \endgroup}
\def\mn@eprint#1#2{\mn@eprint@#1:#2::\@nil}
\def\mn@eprint@arXiv#1{\href {http://arxiv.org/abs/#1} {{\tt arXiv:#1}}}
\def\mn@eprint@dblp#1{\href {http://dblp.uni-trier.de/rec/bibtex/#1.xml}
  {dblp:#1}}
\def\mn@eprint@#1:#2:#3:#4\@nil{\def\@tempa {#1}\def\@tempb {#2}\def\@tempc
  {#3}\ifx \@tempc \@empty \let \@tempc \@tempb \let \@tempb \@tempa \fi \ifx
  \@tempb \@empty \def\@tempb {arXiv}\fi \@ifundefined
  {mn@eprint@\@tempb}{\@tempb:\@tempc}{\expandafter \expandafter \csname
  mn@eprint@\@tempb\endcsname \expandafter{\@tempc}}}

\bibitem[\protect\citeauthoryear{{Balogh} et~al.,}{{Balogh}
  et~al.}{2004}]{2004MNRAS.348.1355B}
{Balogh} M.,  et~al., 2004, \mn@doi [\mnras]
  {10.1111/j.1365-2966.2004.07453.x}, \href
  {https://ui.adsabs.harvard.edu/abs/2004MNRAS.348.1355B} {348, 1355}

\bibitem[\protect\citeauthoryear{Barnes \& Webster}{Barnes \&
  Webster}{2001a}]{10.1046/j.1365-8711.2001.04273.x}
Barnes D.~G.,  Webster R.~L.,  2001a, \mn@doi [\mnras]
  {10.1046/j.1365-8711.2001.04273.x}, 324, 859

\bibitem[\protect\citeauthoryear{{Barnes} \& {Webster}}{{Barnes} \&
  {Webster}}{2001b}]{2001MNRAS.324..859B}
{Barnes} D.~G.,  {Webster} R.~L.,  2001b, \mn@doi [\mnras]
  {10.1046/j.1365-8711.2001.04273.x}, \href
  {https://ui.adsabs.harvard.edu/abs/2001MNRAS.324..859B} {324, 859}

\bibitem[\protect\citeauthoryear{Barnes et~al.,}{Barnes
  et~al.}{2001}]{10.1046/j.1365-8711.2001.04102.x}
Barnes D.~G.,  et~al., 2001, \mn@doi [\mnras]
  {10.1046/j.1365-8711.2001.04102.x}, 322, 486

\bibitem[\protect\citeauthoryear{{Bekki}, {Koribalski}, {Ryder}  \&
  {Couch}}{{Bekki} et~al.}{2005a}]{2005MNRAS.357L..21B}
{Bekki} K.,  {Koribalski} B.~S.,  {Ryder} S.~D.,   {Couch} W.~J.,  2005a,
  \mn@doi [\mnras] {10.1111/j.1745-3933.2005.08625.x}, \href
  {https://ui.adsabs.harvard.edu/abs/2005MNRAS.357L..21B} {357, L21}

\bibitem[\protect\citeauthoryear{{Bekki}, {Koribalski}  \& {Kilborn}}{{Bekki}
  et~al.}{2005b}]{2005MNRAS.363L..21B}
{Bekki} K.,  {Koribalski} B.~S.,   {Kilborn} V.~A.,  2005b, \mn@doi [\mnras]
  {10.1111/j.1745-3933.2005.00076.x}, \href
  {https://ui.adsabs.harvard.edu/abs/2005MNRAS.363L..21B} {363, L21}

\bibitem[\protect\citeauthoryear{{Bigiel} \& {Blitz}}{{Bigiel} \&
  {Blitz}}{2012}]{2012ApJ...756..183B}
{Bigiel} F.,  {Blitz} L.,  2012, \mn@doi [\apj] {10.1088/0004-637X/756/2/183},
  \href {https://ui.adsabs.harvard.edu/abs/2012ApJ...756..183B} {756, 183}

\bibitem[\protect\citeauthoryear{{Bournaud} \& {Duc}}{{Bournaud} \&
  {Duc}}{2006}]{2006A&A...456..481B}
{Bournaud} F.,  {Duc} P.~A.,  2006, \mn@doi [\aap]
  {10.1051/0004-6361:20065248}, \href
  {https://ui.adsabs.harvard.edu/abs/2006A&A...456..481B} {456, 481}

\bibitem[\protect\citeauthoryear{{Bournaud} et~al.,}{{Bournaud}
  et~al.}{2007}]{2007Sci...316.1166B}
{Bournaud} F.,  et~al., 2007, \mn@doi [Science] {10.1126/science.1142114},
  \href {https://ui.adsabs.harvard.edu/abs/2007Sci...316.1166B} {316, 1166}

\bibitem[\protect\citeauthoryear{{Camilo} et~al.,}{{Camilo}
  et~al.}{2018}]{2018ApJ...856..180C}
{Camilo} F.,  et~al., 2018, \mn@doi [\apj] {10.3847/1538-4357/aab35a}, \href
  {https://ui.adsabs.harvard.edu/abs/2018ApJ...856..180C} {856, 180}

\bibitem[\protect\citeauthoryear{{Chung}, {van Gorkom}, {Kenney}, {Crowl}  \&
  {Vollmer}}{{Chung} et~al.}{2009}]{2009AJ....138.1741C}
{Chung} A.,  {van Gorkom} J.~H.,  {Kenney} J. D.~P.,  {Crowl} H.,   {Vollmer}
  B.,  2009, \mn@doi [\aj] {10.1088/0004-6256/138/6/1741}, \href
  {https://ui.adsabs.harvard.edu/abs/2009AJ....138.1741C} {138, 1741}

\bibitem[\protect\citeauthoryear{{Dey} et~al.,}{{Dey}
  et~al.}{2019}]{2019AJ....157..168D}
{Dey} A.,  et~al., 2019, \mn@doi [\aj] {10.3847/1538-3881/ab089d}, \href
  {https://ui.adsabs.harvard.edu/abs/2019AJ....157..168D} {157, 168}

\bibitem[\protect\citeauthoryear{{Dressler}}{{Dressler}}{1980}]{1980ApJ...236..351D}
{Dressler} A.,  1980, \mn@doi [\apj] {10.1086/157753}, \href
  {https://ui.adsabs.harvard.edu/abs/1980ApJ...236..351D} {236, 351}

\bibitem[\protect\citeauthoryear{{Dressler}}{{Dressler}}{2004}]{2004ogci.conf..341D}
{Dressler} A.,  2004, in {Diaferio} A.,  ed., IAU Colloq. 195: Outskirts of
  Galaxy Clusters: Intense Life in the Suburbs. pp 341--346,
  \mn@doi{10.1017/S1743921304000730}

\bibitem[\protect\citeauthoryear{{English}, {Koribalski}, {Bland-Hawthorn},
  {Freeman}  \& {McCain}}{{English} et~al.}{2010}]{2010AJ....139..102E}
{English} J.,  {Koribalski} B.,  {Bland-Hawthorn} J.,  {Freeman} K.~C.,
  {McCain} C.~F.,  2010, \mn@doi [\aj] {10.1088/0004-6256/139/1/102}, \href
  {https://ui.adsabs.harvard.edu/abs/2010AJ....139..102E} {139, 102}

\bibitem[\protect\citeauthoryear{{Garcia}}{{Garcia}}{1995}]{1995A&A...297...56G}
{Garcia} A.~M.,  1995, \aap, \href
  {https://ui.adsabs.harvard.edu/abs/1995A&A...297...56G} {297, 56}

\bibitem[\protect\citeauthoryear{{Gunn} \& {Gott}}{{Gunn} \&
  {Gott}}{1972}]{1972ApJ...176....1G}
{Gunn} J.~E.,  {Gott} J.~Richard I.,  1972, \mn@doi [\apj] {10.1086/151605},
  \href {https://ui.adsabs.harvard.edu/abs/1972ApJ...176....1G} {176, 1}

\bibitem[\protect\citeauthoryear{{Haynes} et~al.,}{{Haynes}
  et~al.}{2018}]{2018ApJ...861...49H}
{Haynes} M.~P.,  et~al., 2018, \mn@doi [\apj] {10.3847/1538-4357/aac956}, \href
  {https://ui.adsabs.harvard.edu/abs/2018ApJ...861...49H} {861, 49}

\bibitem[\protect\citeauthoryear{{Hess} \& {Wilcots}}{{Hess} \&
  {Wilcots}}{2013}]{2013AJ....146..124H}
{Hess} K.~M.,  {Wilcots} E.~M.,  2013, \mn@doi [\aj]
  {10.1088/0004-6256/146/5/124}, \href
  {https://ui.adsabs.harvard.edu/abs/2013AJ....146..124H} {146, 124}

\bibitem[\protect\citeauthoryear{{Hess}, {Cluver}, {Yahya}, {Leisman}, {Serra},
  {Lucero}, {Passmoor}  \& {Carignan}}{{Hess}
  et~al.}{2017}]{2017MNRAS.464..957H}
{Hess} K.~M.,  {Cluver} M.~E.,  {Yahya} S.,  {Leisman} L.,  {Serra} P.,
  {Lucero} D.~M.,  {Passmoor} S.~S.,   {Carignan} C.,  2017, \mn@doi [\mnras]
  {10.1093/mnras/stw2338}, \href
  {https://ui.adsabs.harvard.edu/abs/2017MNRAS.464..957H} {464, 957}

\bibitem[\protect\citeauthoryear{{Jones} et~al.,}{{Jones}
  et~al.}{2018}]{2018A&A...609A..17J}
{Jones} M.~G.,  et~al., 2018, \mn@doi [\aap] {10.1051/0004-6361/201731448},
  \href {https://ui.adsabs.harvard.edu/abs/2018A&A...609A..17J} {609, A17}

\bibitem[\protect\citeauthoryear{{Jones} et~al.,}{{Jones}
  et~al.}{2019}]{2019A&A...632A..78J}
{Jones} M.~G.,  et~al., 2019, \mn@doi [\aap] {10.1051/0004-6361/201936349},
  \href {https://ui.adsabs.harvard.edu/abs/2019A&A...632A..78J} {632, A78}

\bibitem[\protect\citeauthoryear{{J{\'o}zsa} et~al.,}{{J{\'o}zsa}
  et~al.}{2020a}]{2020ascl.soft06014J}
{J{\'o}zsa} G. I.~G.,  et~al., 2020a, {CARACal: Containerized Automated Radio
  Astronomy Calibration pipeline} (\mn@eprint {ascl} {2006.014})

\bibitem[\protect\citeauthoryear{{J\'ozsa} et~al.,}{{J\'ozsa}
  et~al.}{2020b}]{Jozsa2020}
{J\'ozsa} G.~I.~G.,  et~al., 2020b, in {Pizzo} R.,  {Deul} E.,  {Mol} J.-D.,
  {de Plaa} J.,   {Verkouter} H.,  eds,  ASP Conf. Ser. Vol. 527, ADASS XXIX.
  San Francisco, pp 635--638

\bibitem[\protect\citeauthoryear{{Kenyon}, {Smirnov}, {Grobler}  \&
  {Perkins}}{{Kenyon} et~al.}{2018}]{2018MNRAS.478.2399K}
{Kenyon} J.~S.,  {Smirnov} O.~M.,  {Grobler} T.~L.,   {Perkins} S.~J.,  2018,
  \mn@doi [\mnras] {10.1093/mnras/sty1221}, \href
  {https://ui.adsabs.harvard.edu/abs/2018MNRAS.478.2399K} {478, 2399}

\bibitem[\protect\citeauthoryear{{Kilborn}, {Koribalski}, {Forbes}, {Barnes}
  \& {Musgrave}}{{Kilborn} et~al.}{2005}]{2005MNRAS.356...77K}
{Kilborn} V.~A.,  {Koribalski} B.~S.,  {Forbes} D.~A.,  {Barnes} D.~G.,
  {Musgrave} R.~C.,  2005, \mn@doi [\mnras] {10.1111/j.1365-2966.2004.08450.x},
  \href {https://ui.adsabs.harvard.edu/abs/2005MNRAS.356...77K} {356, 77}

\bibitem[\protect\citeauthoryear{Kleiner et~al.,}{Kleiner
  et~al.}{2019}]{10.1093/mnras/stz2063}
Kleiner D.,  et~al., 2019, \mn@doi [\mnras] {10.1093/mnras/stz2063}, 488, 5352

\bibitem[\protect\citeauthoryear{{Kleiner} et~al.,}{{Kleiner}
  et~al.}{2021}]{2021arXiv210110347K}
{Kleiner} D.,  et~al., 2021, arXiv e-prints, \href
  {https://ui.adsabs.harvard.edu/abs/2021arXiv210110347K} {p. arXiv:2101.10347}

\bibitem[\protect\citeauthoryear{{Knobel}, {Lilly}, {Woo}  \&
  {Kova{\v{c}}}}{{Knobel} et~al.}{2015}]{2015ApJ...800...24K}
{Knobel} C.,  {Lilly} S.~J.,  {Woo} J.,   {Kova{\v{c}}} K.,  2015, \mn@doi
  [\apj] {10.1088/0004-637X/800/1/24}, \href
  {https://ui.adsabs.harvard.edu/abs/2015ApJ...800...24K} {800, 24}

\bibitem[\protect\citeauthoryear{{Koribalski}}{{Koribalski}}{2012}]{2012PASA...29..359K}
{Koribalski} B.~S.,  2012, \mn@doi [\pasa] {10.1071/AS12030}, \href
  {https://ui.adsabs.harvard.edu/abs/2012PASA...29..359K} {29, 359}

\bibitem[\protect\citeauthoryear{{Koribalski}}{{Koribalski}}{2020}]{2020arXiv200207312K}
{Koribalski} B.~S.,  2020, arXiv e-prints, \href
  {https://ui.adsabs.harvard.edu/abs/2020arXiv200207312K} {p. arXiv:2002.07312}

\bibitem[\protect\citeauthoryear{{Koribalski} \& {Dickey}}{{Koribalski} \&
  {Dickey}}{2004}]{2004MNRAS.348.1255K}
{Koribalski} B.,  {Dickey} J.~M.,  2004, \mn@doi [\mnras]
  {10.1111/j.1365-2966.2004.07444.x}, \href
  {https://ui.adsabs.harvard.edu/abs/2004MNRAS.348.1255K} {348, 1255}

\bibitem[\protect\citeauthoryear{{Koribalski} et~al.,}{{Koribalski}
  et~al.}{2004}]{2004AJ....128...16K}
{Koribalski} B.~S.,  et~al., 2004, \mn@doi [\aj] {10.1086/421744}, \href
  {https://ui.adsabs.harvard.edu/abs/2004AJ....128...16K} {128, 16}

\bibitem[\protect\citeauthoryear{{Koribalski} et~al.,}{{Koribalski}
  et~al.}{2018}]{2018MNRAS.478.1611K}
{Koribalski} B.~S.,  et~al., 2018, \mn@doi [\mnras] {10.1093/mnras/sty479},
  \href {https://ui.adsabs.harvard.edu/abs/2018MNRAS.478.1611K} {478, 1611}

\bibitem[\protect\citeauthoryear{{Koribalski} et~al.,}{{Koribalski}
  et~al.}{2020}]{2020Ap&SS.365..118K}
{Koribalski} B.~S.,  et~al., 2020, \mn@doi [\apss]
  {10.1007/s10509-020-03831-4}, \href
  {https://ui.adsabs.harvard.edu/abs/2020Ap&SS.365..118K} {365, 118}

\bibitem[\protect\citeauthoryear{{Kraft}, {Jones}, {Nulsen}  \&
  {Hardcastle}}{{Kraft} et~al.}{2006}]{2006ApJ...640..762K}
{Kraft} R.~P.,  {Jones} C.,  {Nulsen} P.~E.~J.,   {Hardcastle} M.~J.,  2006,
  \mn@doi [\apj] {10.1086/500123}, \href
  {https://ui.adsabs.harvard.edu/abs/2006ApJ...640..762K} {640, 762}

\bibitem[\protect\citeauthoryear{Lee-Waddell et~al.,}{Lee-Waddell
  et~al.}{2016}]{10.1093/mnras/stw1162}
Lee-Waddell K.,  et~al., 2016, \mn@doi [\mnras] {10.1093/mnras/stw1162}, 460,
  2945

\bibitem[\protect\citeauthoryear{{Lee-Waddell} et~al.,}{{Lee-Waddell}
  et~al.}{2019}]{2019MNRAS.487.5248L}
{Lee-Waddell} K.,  et~al., 2019, \mn@doi [\mnras] {10.1093/mnras/stz017}, \href
  {https://ui.adsabs.harvard.edu/abs/2019MNRAS.487.5248L} {487, 5248}

\bibitem[\protect\citeauthoryear{{Lisenfeld} et~al.,}{{Lisenfeld}
  et~al.}{2011}]{2011A&A...534A.102L}
{Lisenfeld} U.,  et~al., 2011, \mn@doi [\aap] {10.1051/0004-6361/201117056},
  \href {https://ui.adsabs.harvard.edu/abs/2011A&A...534A.102L} {534, A102}

\bibitem[\protect\citeauthoryear{{Makarov}, {Prugniel}, {Terekhova}, {Courtois}
   \& {Vauglin}}{{Makarov} et~al.}{2014}]{2014A&A...570A..13M}
{Makarov} D.,  {Prugniel} P.,  {Terekhova} N.,  {Courtois} H.,   {Vauglin} I.,
  2014, \mn@doi [\aap] {10.1051/0004-6361/201423496}, \href
  {https://ui.adsabs.harvard.edu/abs/2014A&A...570A..13M} {570, A13}

\bibitem[\protect\citeauthoryear{{McMullin}, {Waters}, {Schiebel}, {Young}  \&
  {Golap}}{{McMullin} et~al.}{2007}]{2007ASPC..376..127M}
{McMullin} J.~P.,  {Waters} B.,  {Schiebel} D.,  {Young} W.,   {Golap} K.,
  2007, {CASA Architecture and Applications}.
p.~127

\bibitem[\protect\citeauthoryear{Meyer et~al.,}{Meyer
  et~al.}{2004}]{10.1111/j.1365-2966.2004.07710.x}
Meyer M.~J.,  et~al., 2004, \mn@doi [\mnras]
  {10.1111/j.1365-2966.2004.07710.x}, 350, 1195

\bibitem[\protect\citeauthoryear{Offringa \& Smirnov}{Offringa \&
  Smirnov}{2017}]{10.1093/mnras/stx1547}
Offringa A.~R.,  Smirnov O.,  2017, \mn@doi [\mnras] {10.1093/mnras/stx1547},
  471, 301

\bibitem[\protect\citeauthoryear{{Offringa} et~al.,}{{Offringa}
  et~al.}{2014}]{2014MNRAS.444..606O}
{Offringa} A.~R.,  et~al., 2014, \mn@doi [\mnras] {10.1093/mnras/stu1368},
  \href {https://ui.adsabs.harvard.edu/abs/2014MNRAS.444..606O} {444, 606}

\bibitem[\protect\citeauthoryear{{Pisano}, {Wilcots}  \& {Elmegreen}}{{Pisano}
  et~al.}{2000}]{2000AJ....120..763P}
{Pisano} D.~J.,  {Wilcots} E.~M.,   {Elmegreen} B.~G.,  2000, \mn@doi [\aj]
  {10.1086/301464}, \href
  {https://ui.adsabs.harvard.edu/abs/2000AJ....120..763P} {120, 763}

\bibitem[\protect\citeauthoryear{Porter, Raychaudhury, Pimbblet  \&
  Drinkwater}{Porter et~al.}{2008}]{10.1111/j.1365-2966.2008.13388.x}
Porter S.~C.,  Raychaudhury S.,  Pimbblet K.~A.,   Drinkwater M.~J.,  2008,
  \mn@doi [\mnras] {10.1111/j.1365-2966.2008.13388.x}, 388, 1152

\bibitem[\protect\citeauthoryear{{Robotham} et~al.,}{{Robotham}
  et~al.}{2011}]{2011MNRAS.416.2640R}
{Robotham} A.~S.~G.,  et~al., 2011, \mn@doi [\mnras]
  {10.1111/j.1365-2966.2011.19217.x}, \href
  {https://ui.adsabs.harvard.edu/abs/2011MNRAS.416.2640R} {416, 2640}

\bibitem[\protect\citeauthoryear{{Ryder} et~al.,}{{Ryder}
  et~al.}{2001}]{2001ApJ...555..232R}
{Ryder} S.~D.,  et~al., 2001, \mn@doi [\apj] {10.1086/321453}, \href
  {https://ui.adsabs.harvard.edu/abs/2001ApJ...555..232R} {555, 232}

\bibitem[\protect\citeauthoryear{Saponara, Koribalski, Benaglia  \&
  FernÃ¡ndez~LÃ³pez}{Saponara et~al.}{2017}]{10.1093/mnras/stx2475}
Saponara J.,  Koribalski B.~S.,  Benaglia P.,   FernÃ¡ndez~LÃ³pez M.,
  2017, \mn@doi [\mnras] {10.1093/mnras/stx2475}, 473, 3358

\bibitem[\protect\citeauthoryear{{Saponara}, {Koribalski}, {Benaglia}  \&
  {Fern{\'a}ndez L{\'o}pez}}{{Saponara} et~al.}{2018}]{2018MNRAS.473.3358S}
{Saponara} J.,  {Koribalski} B.~S.,  {Benaglia} P.,   {Fern{\'a}ndez L{\'o}pez}
  M.,  2018, \mn@doi [\mnras] {10.1093/mnras/stx2475}, \href
  {https://ui.adsabs.harvard.edu/abs/2018MNRAS.473.3358S} {473, 3358}

\bibitem[\protect\citeauthoryear{{Saulder}, {van Kampen}, {Chilingarian},
  {Mieske}  \& {Zeilinger}}{{Saulder} et~al.}{2016}]{2016A&A...596A..14S}
{Saulder} C.,  {van Kampen} E.,  {Chilingarian} I.~V.,  {Mieske} S.,
  {Zeilinger} W.~W.,  2016, \mn@doi [\aap] {10.1051/0004-6361/201526711}, \href
  {https://ui.adsabs.harvard.edu/abs/2016A&A...596A..14S} {596, A14}

\bibitem[\protect\citeauthoryear{{Schr{\"o}der}, {Drinkwater}  \&
  {Richter}}{{Schr{\"o}der} et~al.}{2001}]{2001A&A...376...98S}
{Schr{\"o}der} A.,  {Drinkwater} M.~J.,   {Richter} O.~G.,  2001, \mn@doi
  [\aap] {10.1051/0004-6361:20010997}, \href
  {https://ui.adsabs.harvard.edu/abs/2001A&A...376...98S} {376, 98}

\bibitem[\protect\citeauthoryear{Sengupta \& Balasubramanyam}{Sengupta \&
  Balasubramanyam}{2006}]{10.1111/j.1365-2966.2006.10307.x}
Sengupta C.,  Balasubramanyam R.,  2006, \mn@doi [\mnras]
  {10.1111/j.1365-2966.2006.10307.x}, 369, 360

\bibitem[\protect\citeauthoryear{{Sengupta}, {Balasubramanyam}  \&
  {Dwarakanath}}{{Sengupta} et~al.}{2007}]{2007MNRAS.378..137S}
{Sengupta} C.,  {Balasubramanyam} R.,   {Dwarakanath} K.~S.,  2007, \mn@doi
  [\mnras] {10.1111/j.1365-2966.2007.11748.x}, \href
  {https://ui.adsabs.harvard.edu/abs/2007MNRAS.378..137S} {378, 137}

\bibitem[\protect\citeauthoryear{Serra et~al.,}{Serra
  et~al.}{2012}]{10.1093/mnras/sts033}
Serra P.,  et~al., 2012, \mn@doi [\mnras] {10.1093/mnras/sts033}, 428, 370

\bibitem[\protect\citeauthoryear{Serra et~al.,}{Serra
  et~al.}{2015}]{10.1093/mnras/stv079}
Serra P.,  et~al., 2015, \mn@doi [\mnras] {10.1093/mnras/stv079}, 448, 1922

\bibitem[\protect\citeauthoryear{Seth \& Raychaudhury}{Seth \&
  Raychaudhury}{2020}]{10.1093/mnras/staa1779}
Seth R.,  Raychaudhury S.,  2020, \mn@doi [\mnras] {10.1093/mnras/staa1779},
  497, 466

\bibitem[\protect\citeauthoryear{Solanes, Manrique, Garc{\'{\i}}a-G{\'{o}}mez,
  Gonz{\'{a}}lez-Casado, Giovanelli  \& Haynes}{Solanes
  et~al.}{2001}]{Solanes_2001}
Solanes J.~M.,  Manrique A.,  Garc{\'{\i}}a-G{\'{o}}mez C.,
  Gonz{\'{a}}lez-Casado G.,  Giovanelli R.,   Haynes M.~P.,  2001, \mn@doi
  [\apj] {10.1086/318672}, 548, 97

\bibitem[\protect\citeauthoryear{Spindler et~al.,}{Spindler
  et~al.}{2018}]{10.1093/mnras/sty247}
Spindler A.,  et~al., 2018, \mn@doi [\mnras] {10.1093/mnras/sty247}, 476, 580

\bibitem[\protect\citeauthoryear{{Tully}}{{Tully}}{1987}]{1987ApJ...321..280T}
{Tully} R.~B.,  1987, \mn@doi [\apj] {10.1086/165629}, \href
  {https://ui.adsabs.harvard.edu/abs/1987ApJ...321..280T} {321, 280}

\bibitem[\protect\citeauthoryear{{Verdes-Montenegro}, {Yun}, {Williams},
  {Huchtmeier}, {Del Olmo}  \& {Perea}}{{Verdes-Montenegro}
  et~al.}{2001}]{2001A&A...377..812V}
{Verdes-Montenegro} L.,  {Yun} M.~S.,  {Williams} B.~A.,  {Huchtmeier} W.~K.,
  {Del Olmo} A.,   {Perea} J.,  2001, \mn@doi [\aap]
  {10.1051/0004-6361:20011127}, \href
  {https://ui.adsabs.harvard.edu/abs/2001A&A...377..812V} {377, 812}

\bibitem[\protect\citeauthoryear{{Verdes-Montenegro}, {Sulentic}, {Lisenfeld},
  {Leon}, {Espada}, {Garcia}, {Sabater}  \& {Verley}}{{Verdes-Montenegro}
  et~al.}{2005}]{2005A&A...436..443V}
{Verdes-Montenegro} L.,  {Sulentic} J.,  {Lisenfeld} U.,  {Leon} S.,  {Espada}
  D.,  {Garcia} E.,  {Sabater} J.,   {Verley} S.,  2005, \mn@doi [\aap]
  {10.1051/0004-6361:20042280}, \href
  {https://ui.adsabs.harvard.edu/abs/2005A&A...436..443V} {436, 443}

\bibitem[\protect\citeauthoryear{{Vogt}, {Owen}, {Verdes-Montenegro}  \&
  {Borthakur}}{{Vogt} et~al.}{2016}]{2016ApJ...818..115V}
{Vogt} F. P.~A.,  {Owen} C.~I.,  {Verdes-Montenegro} L.,   {Borthakur} S.,
  2016, \mn@doi [\apj] {10.3847/0004-637X/818/2/115}, \href
  {https://ui.adsabs.harvard.edu/abs/2016ApJ...818..115V} {818, 115}

\bibitem[\protect\citeauthoryear{Zabludoff \& Mulchaey}{Zabludoff \&
  Mulchaey}{1998}]{Zabludoff_1998}
Zabludoff A.~I.,  Mulchaey J.~S.,  1998, \mn@doi [\apj] {10.1086/305355}, 496,
  39

\bibitem[\protect\citeauthoryear{{de Vaucouleurs}, {de Vaucouleurs}, {Corwin},
  {Buta}, {Paturel}  \& {Fouque}}{{de Vaucouleurs}
  et~al.}{1991}]{1991rc3..book.....D}
{de Vaucouleurs} G.,  {de Vaucouleurs} A.,  {Corwin} Herold~G. J.,  {Buta}
  R.~J.,  {Paturel} G.,   {Fouque} P.,  1991, {Third Reference Catalogue of
  Bright Galaxies}

\makeatother
\end{thebibliography}
\end{document}